\documentclass[aps,prd,showpacs,11pt]{revtex4-1}

\usepackage{graphicx,epsfig}
\usepackage{latexsym,amssymb,amsmath,float,url}
\usepackage[T1]{fontenc} 
\usepackage[ansinew]{inputenc}
\usepackage{epstopdf}
\usepackage{slashed}
\usepackage{bm} 
\usepackage{epsf}
  
\newcommand{\rarr}{\rightarrow}

\newcommand{\fbar}{{\bar f}}
\newcommand{\Psibar}{{\bar\Psi}}

\newcommand{\kbar}{{\bar k}}

\newcommand{\hT}{{\hat T}}
\newcommand{\Ghat}{{\hat\Gamma}}
\newcommand{\ubar}{{\overline u}}
\newcommand{\vbar}{{\overline v}}

\newcommand{\X}{{\rm X}}
\newcommand{\Y}{{\rm Y}}
\newcommand{\chibar}{{\overline\chi}}
\newcommand{\ellbar}{{\overline\ell}}

\mathchardef\mhyphen="2D 
\newcommand{\half}{\frac{1}{2}}
\newcommand{\quarter}{\frac{1}{4}}

\newcommand{\beq}[1]{\begin{equation}\label{#1}}
\newcommand{\eeq}{\end{equation}}
\newcommand{\bea}[1]{\begin{eqnarray}\label{#1}}
\newcommand{\eea}{\end{eqnarray}}

\newcommand{\rf}[1]{(\ref{#1})}

\newcommand{\barr}{\begin{array}}
\newcommand{\earr}{\end{array}}
\def\be{\begin{equation}}
\def\ee{\end{equation}}
\def\ba{\begin{eqnarray}}
\def\ea{\end{eqnarray}}
\begin{document}

\title{On the likely dominance of  WIMP annihilation to \boldmath{$ f\fbar+W/Z$} \\
 (and implication for indirect detection)}

\author{Thomas J. Weiler}
\email[Electronic mail: ]{tom.weiler@vanderbilt.edu}
\affiliation{Department of Physics and Astronomy, Vanderbilt University, Nashville, TN 37235, USA}

\begin{abstract}
Arguably, the most popular candidate for Dark Matter (DM) is a massive, stable, Majorana fermion.  
However, annihilation of Majorana DM to two fermions often features a helicity-suppressed $s$-wave rate.
Radiating a gauge boson via electroweak (EW) and electromagnetic (EM)~bremsstrahlung 
removes this $s$-wave suppression.
The main purpose of this talk is to explain in some detail why the branching ratio 
Br($\rarr f\fbar$) is likely suppressed while Br($\rarr f\fbar+W/Z/\gamma $) is not.
In doing so, we investigate the general conditions for $s$-wave suppression and un-suppression
using Fierz transformations and partial wave expansions.
Suppression for the $2\rarr 2$ process is sufficiently severe that the 
EW and EM bremsstrahlung are likely to be the dominant modes of 
gauge-singlet Majorana DM annihilation.
We end this talk with a discussion of the 
challenge presented by space-based data 
for Majorana DM models,
given that the enhanced rate to radiated $W$ and $Z$ gauge bosons and their
dominant decay via hadronic channels tends to produce more antiprotons than are observed.
\end{abstract}



\maketitle

\section{The Storyline}
\label{sec:storyline}
I this talk I summarize the key results of work published in collaboration with N. Bell, J. Dent, and T. Jacques,
and later with A. Galea and L. Krauss as well, on the decay dynamics of Majorana fermion 
Dark Matter (DM) ~\cite{Bell:2008ey,Bell:2010ei,Bell:2011eu,Bell:2011if}.  
Any citations to topics in this talk should likely target one or more of these original references.

Weakly coupled particles with a weak scale mass decouple from thermal equilibrium in the Early Universe 
at the right moment to yield the fractional DM energy density which we observe today, $\Omega_{\rm DM}\sim 0.24$.
The so-called ``WIMP miracle'' (WIMP, for Weakly Interacting Massive Particle) is that 
$\sigma_{\rm ann}\sim (\frac{\alpha_{\rm weak}}{4\pi})^2/M_{\rm WIMP}^2 \sim$~picobarn$\times(\frac{200\,{\rm GeV}}{M_{\rm WIMP}})^2$
yields the critical thermally-averaged annihilation ``rate''
$\langle{\rm v}\sigma\rangle \sim 3\times 10^{-26}{\rm cm}^3/s$
to effect a DM abundance in agreement with what is observed today.
Here, v~is the DM M\"oller velocity~\cite{note1}
(in units of $c$) 
and $\sigma_{\rm ann}$ is the DM annihilation cross section.

Among the many candidates for Dark Matter (DM), the most common particle-type is the Majorana fermion.
A splendid example is the neutralino ($\chi$)  of Supersymmetry (SUSY) models.
An $R$-parity can be defined under which the Lightest Supersymmetric Particle (LSP) is absolutely stable.
In many SUSY models, the LSP is the neutralino.
Being a partner to a massless gauge-boson in the unbroken SUSY theory, the neutralino has two independent spin states, not four.  
Thus, $\chi$ is a Majorana fermion and the two annihilating $\chi$'s are identical particles.
In this talk we focus not on iso-vector DM like the neutralino of SUSY, but rather on gauge-singlet Majorana DM.

Majorana-ness means that in the annihilation process to a fermion pair, $\chi\,\chi\rarr f\fbar$, 
the two initial state $\chi$'s contribute two amplitudes, the usual one and a crossed-$\chi$ diagram with a relative minus sign.
The two diagrams result from  Figs.~\rf{fig:feyngraphs_ace} and \rf{fig:feyngraphs_bdf} when the external gauge boson lines are removed.
The process is mediated by a $t$- and $u$-channel exchange of a virtual intermediate particle which we label $\eta$.
When the fermion currents (connected fermion lines) of these two graphs 
are Fierz rearranged into ``charge retention'' order (a $\chi$ line and a light fermion line),
the result is an axial vector coupling in the $s$-channel, plus corrections proportional to the $\chi$'s velocity, v.
%
%
If has been known for many years~\cite{Haim1983} that the spin/orbital~angular momentum of the 
$s$-channel axial-vector coupling requires a helicity flip of one of the produced fermions in the 
$L=0$ (so-called $s$-wave) amplitude.
Thus, the $s$-wave contribution to the rate is suppressed as $(m_f/M_\chi)^2\ll 1$.

The $L=1$ (so-called $p$-wave) amplitude does not require a helicity flip, and so is not suppressed by the mass ratio.
However, on general grounds, the contribution of the $L^{th}$ partial wave to the rate goes dominantly as ${\rm v}^{2L}$~\cite{note2},
and so the $L=1$ $p$-wave is suppressed by $\langle{\rm v}^2\rangle$.  
In parametrized form, the rate can be written as $\langle{\rm v}\sigma\rangle = a +\langle{\rm v}^2\rangle\,b+\cdots$,
where the constant $a$ comes from s-wave annihilation, while the
velocity suppressed $\langle{\rm v}^2\rangle\,b$ term receives both s-wave and p-wave contributions.  

From standard statistical mechanics arguments, one has the relation
$\langle{\rm v}^2\rangle\sim 6\frac{T_{\rm dec}}{M_\chi} (1+\frac{3}{2}\frac{T_{\rm dec}}{M_\chi})$,
where $T_{\rm dec}$ is the decoupling temperature of the DM particle.
Simulations reveal that decoupling occurs at $T_{\rm dec}/M_\chi \sim 1/20\ {\rm to\ }1/50$;
the resulting $\langle{\rm v}^2\rangle$ is therefore
$\sim 0.1\ {\rm to\ }0.3$, which is semi-relativistic.
Consequently, the $p$-wave contribution in  $a+\langle{\rm v}^2\rangle\,b\sim 3\times 10^{-26}$cm$^3$/s 
is velocity-suppressed by at most a factor of~0.1 at DM decoupling in the Early Universe\@.
Today, v$\sim 300\ {\rm km/s}\sim 10^{-3}c$ 
in galactic halos, so the p-wave contribution is highly
suppressed by $\sim 10^{-6}$ and only the s-wave contribution is expected to be significant.  
However, as we have mentioned, in many DM models the s-wave annihilation into a fermion pair $\chi\chi \rarr f\fbar$ 
is helicity suppressed by a factor $(m_f/M_\chi)^2$ 
(so that among the SM $f\fbar$ states, only $\rightarrow \bar{t}t$ modes remain of potential interest~\cite{marc}).
Alternatively, Majorana DM requires artificially large ``boost'' (a new variety of fudge?) factors today 
to produce observable annihilation signals for indirect DM detection.  
These boost factors may be astrophysical, such as sub-clustering of the DM,
or they may arise from particle physics, like the ``Sommerfeld enhancement'' of annihilation if the DM pair 
enjoys certain resonance behavior.
Although not ruled out, these boost factors seem to be a contrivance.

These two suppressions, mass/helicity for the $s$-wave and velocity for the $p$-wave, are quite general.
The helicity-suppression of the $s$-wave amplitude applies not only to neutralino DM $\rarr f\fbar$, 
but to all Majorana fermion DM.
The velocity-suppressed $p$-wave applies to all DM, Majorana or otherwise.

It is becoming increasingly appreciated that if the light fermion pair is produced in association with a gauge boson,
then the spins of the final sate can match the $s$-wave angular momentum requirement without a helicity flip.
Thus, there is no $s$-wave mass-suppression factor for the $2\rarr 3$ process $\chi\chi\rarr f\fbar$+~gauge-boson.
This unsuppressed $s$-wave was first calculated in 1989 for the photon~bremsstrahlung reaction 
$\chi\chi\rarr  f\fbar\gamma$~\cite{gamma1,gamma2},
where the photon is radiated from one of the external particle legs
(final state radiation, FSR) or from a the virtual mediator particle $\eta$
(internal bremsstrahlung, IB).  Gauge invariance requires inclusion of both contributions.
On the face of it, the radiative rate is down by the usual QED coupling factor of
$\alpha/4\pi\sim 10^{-3}$.  However, and significantly, photon
bremsstrahlung can lift the helicity suppression of the $s$-wave
process, which more than compensates for the extra coupling factor.
This un-suppression due to photon bremsstrahlung 
has been well examined in recent publications~~\cite{gamma3,gamma4,gamma5,gamma6}. 
(If the dark matter annihilates to colored fermions, radiation of a gluon would also lift the helicity suppression.)

\begin{figure*}[t]
\includegraphics[height=0.18\textheight,width=0.32\columnwidth]{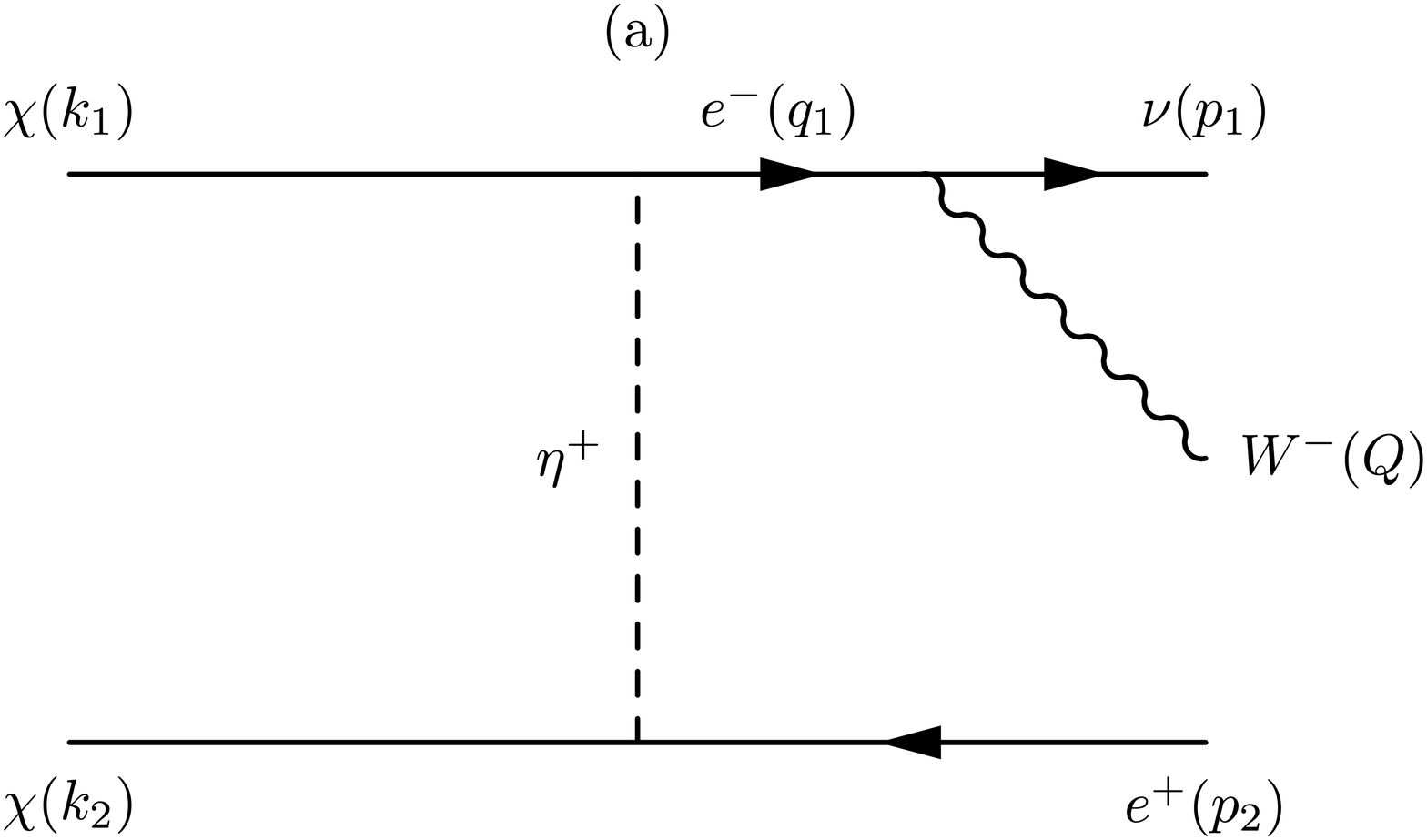}
\includegraphics[height=0.18\textheight,width=0.32\columnwidth]{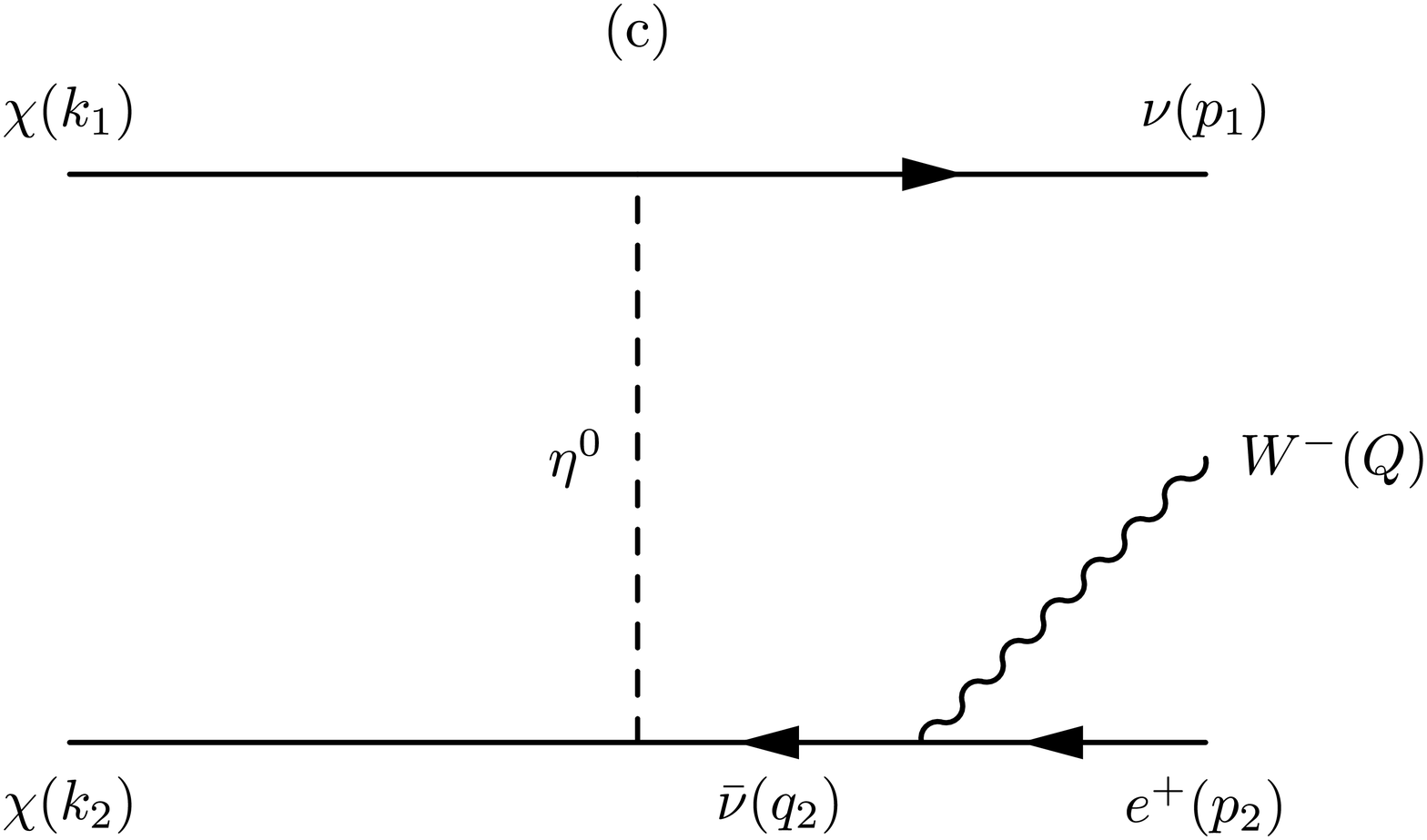}
\includegraphics[height=0.18\textheight,width=0.32\columnwidth]{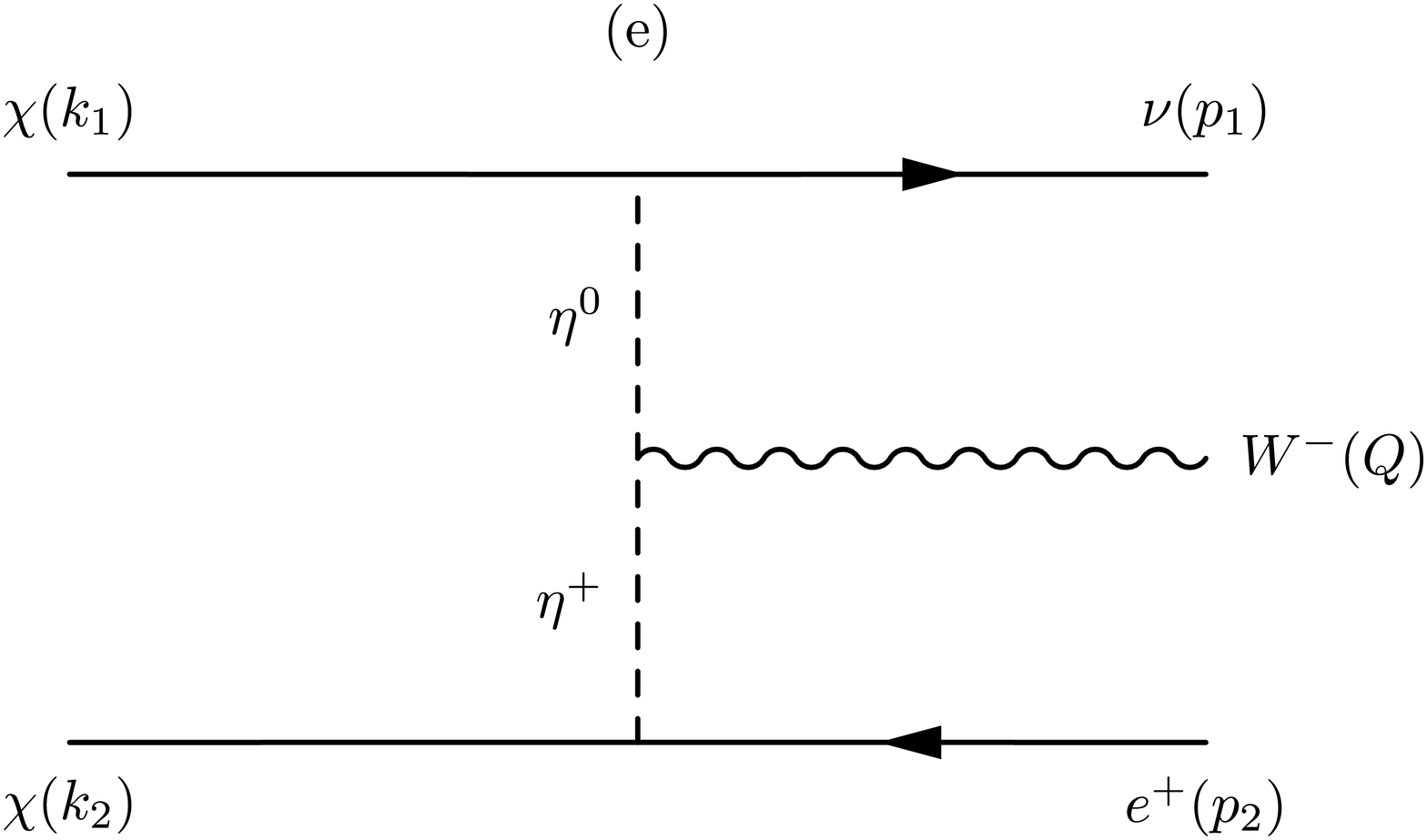}
\caption{
\label{fig:feyngraphs_ace} 
The $t$-channel Feynman diagrams for $\chi\chi\rightarrow e^+\nu W^-$.  
All fermion momenta in the diagrams flow with the arrow except $p_2$ and $q_2$, 
with $q_1=p_1+Q$, $q_2=p_2+Q$.}
\end{figure*}

\begin{figure*}[t]
\includegraphics[height=0.18\textheight,width=0.32\columnwidth]{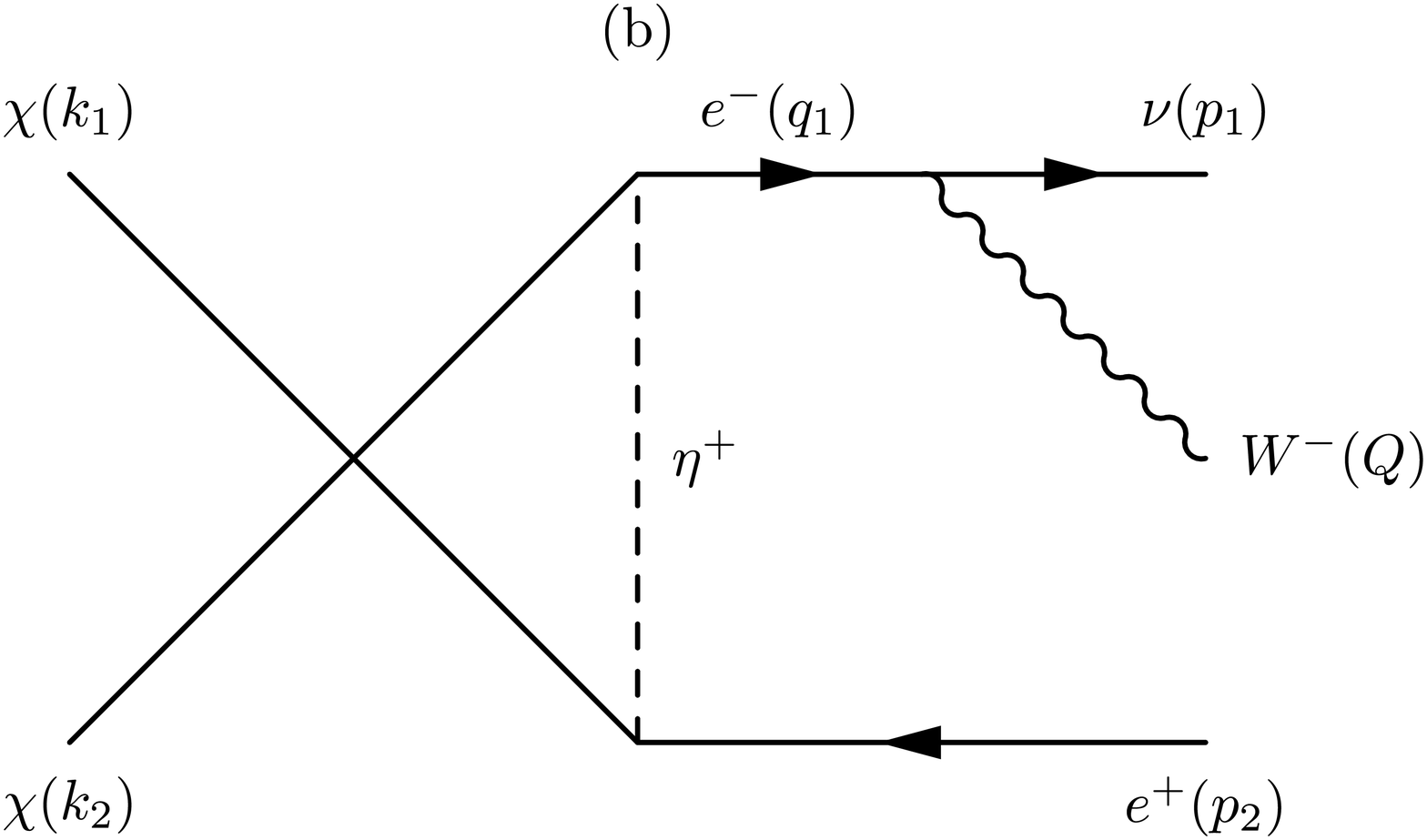}
\includegraphics[height=0.18\textheight,width=0.32\columnwidth]{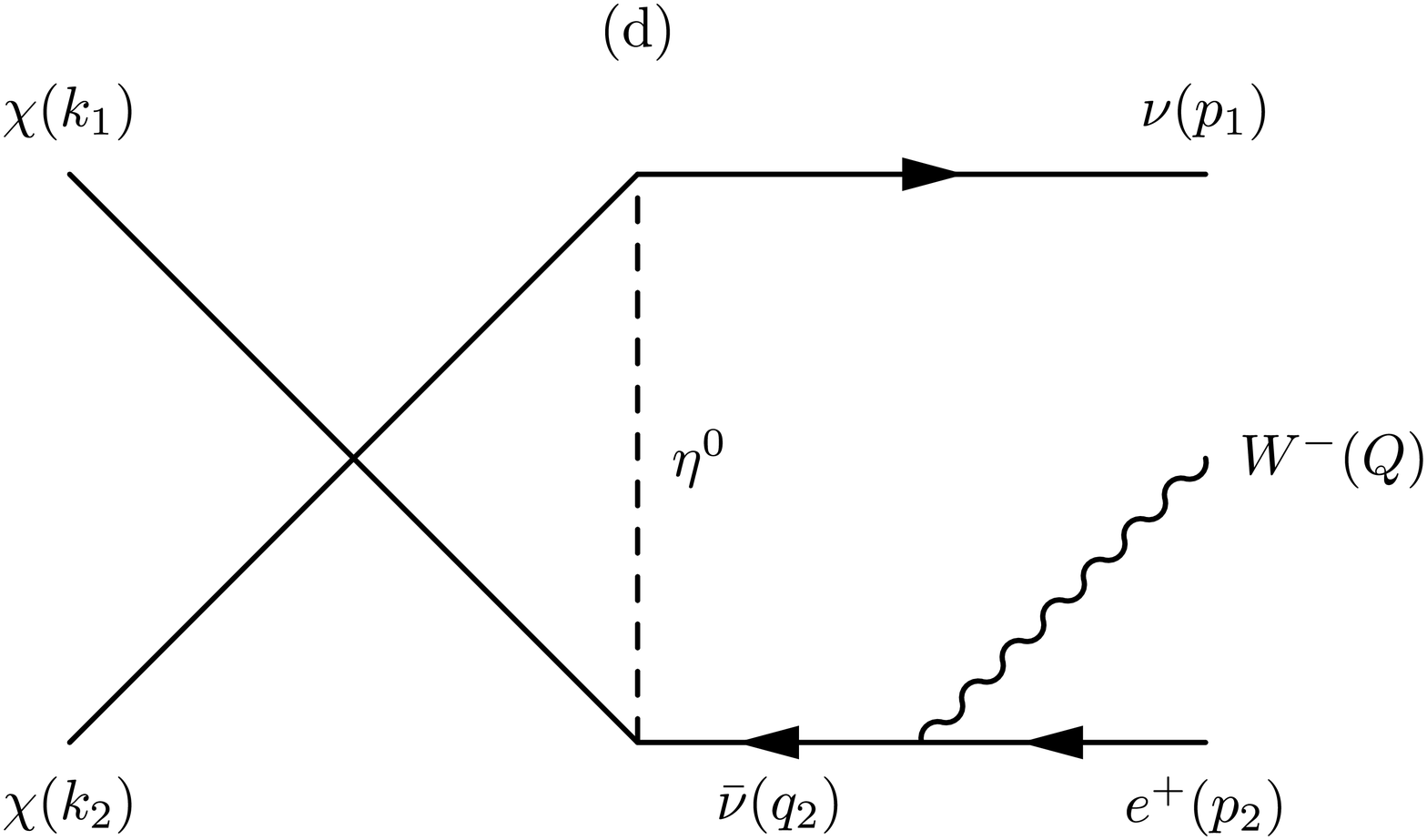}
\includegraphics[height=0.18\textheight,width=0.32\columnwidth]{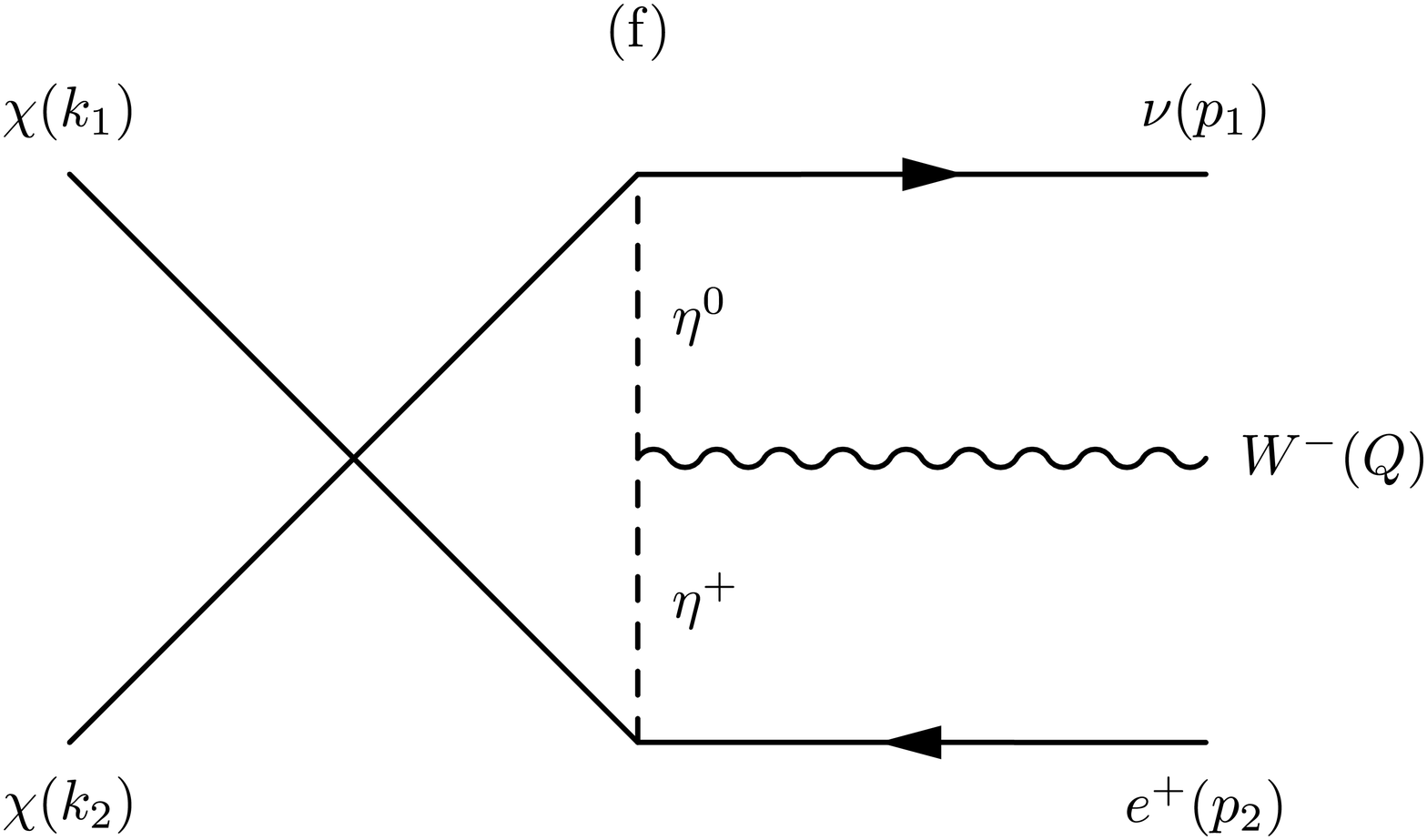}
\caption{
\label{fig:feyngraphs_bdf} 
The $u$-channel Feynman diagrams. }
\end{figure*}

The importance of electroweak radiative corrections to dark matter
annihilation was recognized more recently.
Electroweak bremsstrahlung was investigated first in the context of cosmic 
rays~\cite{Berezinsky:2002hq,Kachelriess:2007aj,Bell:2008ey,Dent:2008qy,Ciafaloni:2010qr,Kachelriess:2009zy,Ciafaloni:2010ti},
and the possibility of  $W/Z$~bremsstrahlung to lift initial-state velocity and 
final-state helicity suppressions was alluded to in~Refs.\cite{Bell:2008ey,Kachelriess:2009zy} but not explored.
We recently noted the unsuppressed $s$-wave in the context of 
$\chi\chi\rarr f\fbar+W/Z$~\cite{Bell:2010ei}.
We had anticipated this result in our earlier paper~\cite{Bell:2008ey} with the statement
``It seems likely to us that this $2\rarr 3$~rate [i.e., EW bremsstrahlung] for the radiatively
corrected $s$-channel axial-vector exchange will exceed the
helicity-suppressed $2\rarr 2$~rate by
$\frac{\alpha}{4\pi}(\frac{M_W}{m_f})^2$, which is many orders of magnitude.''
In~\cite{Bell:2011eu}, we calculated some signature channels for indirect detection of DM~$\chi$'s.
In particular, we made the point that although the $\chi\chi\rarr 2$ process can be tuned by fiat to suppress antiproton production, 
the $2\rarr 3$ EW process will necessarily produce an antiproton signal via the production and hadronic decay of the $W$ and $Z$\@.
As a result, the experimental upper limit on the cosmic antiproton flux provides a meaningful constraint on Majorana DM~\cite{Bell:2011eu},
which we present and discuss at the end of this talk.
In~\cite{Bell:2011if}, we provided further details on the unsuppressed rate of $W/Z$-bremsstrahlung,
and we showed the conditions under which this process becomes the dominant
annihilation channel.
Following that paper, we examined the signal-to-background for the crossed process 
parton+parton$\rarr \chi\chi$+mono-$Z$, 
which is relevant for the DM search at the LHC~\cite{Bell:2012rg}.
The dominance of the helicity-unsuppressed $f\fbar W/Z$~channel has also been elaborated 
upon in Refs.~\cite{Ciafaloni:2011sa,Ciafaloni:2011gv}.
The diagrams contributing to EW bremsstrahlung are shown in Fig.~\ref{fig:feyngraphs_ace} for the $t$-channel,
and Fig.~\ref{fig:feyngraphs_bdf} for the $u$-channel.


There are a few important distinctions between electromagnetic (EM)
and electroweak (EW) bremsstrahlung. An obvious one is that EM bremsstrahlung
produces just photons, whereas EW bremsstrahlung and subsequent decay
of the $W$ or $Z$ leads to leptons (including positrons), hadrons (including antiprotons) 
and gamma rays, offering
correlated ``multi-messenger'' signals for indirect dark matter searches.
This is an important result for future DM searches.
Another distinction between EW and EM~bremsstrahlung is that in the former 
the massive $W/Z$'s have a longitudinal mode not available to the photon.
I will allude to recent work~\cite{Garny:2011cj} on the longitudinal mode later in this talk.

On the experimental astrophysics front, the PAMELA balloon has
observed a sharp excess in the $e^+/(e^-$+$\,e^+)$ fraction at energies
beyond approximately 10 GeV~\cite{PamelaPositrons}, without a
corresponding excess in the antiproton/proton
data~\cite{PamelaAntiprotons}, while the Fermi satellite~\cite{Fermi1} 
and ground-based HESS~\cite{Aharonian:2009ah} have
reported more modest excesses in the $(e^-$+$\,e^+)$ flux at energies of
order 1 TeV.  
New astrophysical sources are thought to ultimately be the mechanism behind the positron data~(see, e.g.,~\cite{Serpico:2011wg}).
However, DM annihilation in the Galactic halo has been advanced as an alternative explanation for the positron excess.
Many of the popular models proposed to explain the positron excess invoke Majorana DM.  
%
%
But as we argued above and develop below (by means of a Fierz transformation of amplitudes into $s$-channel partial waves),
annihilation of Majorana DM to a SM fermion pair is helicity-suppressed in the $s$-wave and velocity-suppressed in the $p$-wave.
We will also argue below that if DM annihilation were ``boosted'' to reproduce the PAMELA positron signal,
then the DM could not be Majorana fermions lest the same annihilation overproduce an antiproton signal.

In the rest of this talk, I will detail the story of the suppressed Majorana DM annihilation channel to $f\fbar$,
and the unsuppressed channel to $f\fbar+W/Z$\@.
I will sometimes denote the light fermions as leptons $\ell$ and $\ellbar$, 
since the leptophilic model of DM annihilation is quite popular.

For $M_\chi \gg M_W$, a mode that dominates neutralino annihilation in SUSY is 
$\chi\chi\rarr W^+\,W^-$.
However, the direct coupling of SU(2)-singlet DM to the SU(2)-triplet $W/Z$ 
is precluded by weak~isospin addition, unless one introduces a new iso-triplet fermion.
I introduce no such particle, and I assume that the Majorana DM is gauge-singlet,
in which case the $\chi\chi\rarr W^+\,W^-$ mode is very suppressed.

\section{Helicity and ${\rm v}^2$ Suppressions Exposed}
\label{sec:suppressions}
In this section, we first present the needed pieces of machinery to elucidate helicity and velocity suppressions, 
and then we put these pieces together to make general conclusions for Majorana DM.
  
\subsection{C, P, and T Constraints on  Majorana Fermions}
\label{subsec:MajoranaCPT}
The discrete symmetries $C$, $P$, and $T$, and angular momentum place 
constraints on fermion pair ${\bar\chi}\chi$ states 
(or the two-fermion final state, or both)~\cite{Bell:2008ey,Bell:2010ei}. 
Two identical fermions comprise a Majorana pair.  A fermion pair can
have total spin $S=1$ in the symmetric state or $S=0$ in the
antisymmetric  state.  The parity of the two-fermion state is
$P=(-)^{L+1}$, where $L$ is the orbital angular momentum of the pair.
This parity formula holds for both Dirac and Majorana pairs.  The
negative intrinsic parity of the pair, independent of the orbital
parity $(-)^L$, is the same for Dirac and Majorana pairs for different
reasons.  In the Dirac case, the $u$ and $v$ spinors (equivalently,
the positive and negative energy states) are independent and have
opposite parity corresponding to the $\pm 1$ eigenvalues of the parity
operator $\gamma^0$.  Reinterpreting the two spinor types 
(the positive and negative energy states) as particle and antiparticle
then leads directly to opposite intrinsic parity for the
particle-antiparticle pair.  In the Majorana case, the fermion has
intrinsic parity $\pm i$, and so the two-particle state has intrinsic
parity $(\pm i)^2=-1$.

On general grounds, the $L^{\rm th}$~partial wave contribution to the
annihilation rate is suppressed as v$^{2L}$, where v is the relative
velocity between the heavy, non-relativistic $\bar\chi \chi$ pair.   
Thus only the $L=0$ partial wave gives an unsuppressed
annihilation rate for DM in today's Universe.  

The general rule for charge-conjugation
(particle-antiparticle exchange) is $C=(-)^{L+S}$.
The origin of this rule is as follows: 
Under particle-antiparticle exchange, the spatial wave
function contributes $(-)^L$, and the spin wave function contributes
(+1) if in the symmetric triplet $S=1$ state, and ($-1$) if in the
antisymmetric $S=0$ singlet state, i.e., $(-)^{S+1}$.  In addition,
there is an overall ($-1$) from anticommutation of the two
particle-creation operators $b^\dag d^\dag$ for the Dirac case, and
$b^\dag b^\dag$ for the Majorana case.
A Majorana pair is even under charge-conjugation, and from
$C=(-)^{L+S}$ one infers that $L$ and $S$ must be either both even,
or both odd for a Majorana pair.  

Consider the $L \le 2$ states.  In spectroscopic notation
$^{(2S+1)}L_J$ and spin-parity notation ($J^{PC}$), the vector $^3
S_1$~$(1^{--}$), $C$-odd axial vector $^1 P_1$~$(1^{+-})$, and
assorted $^3 D_J$~$(J^{--})$ states are all $C$-odd and therefore
disallowed for a Majorana pair.  
The pseudo-scalar $^1 S_0$~($0^{-+})$, scalar $^3
P_0$~($0^{++})$, axial vector $^3 P_1$~($1^{++})$, $C$-even tensor $^3
P_2$~($2^{++})$, and pseudo-tensor $^1 D_2$~($2^{-+}$) are all $C$-even
and therefore allowed.  In particular, the sole $L=0$~state, with no
v$^{2L}$ suppression, is the pseudo-scalar $^1 S_0$~($0^{-+})$.

Incidentally, at threshold, defined by $s=(2M_\chi)^2$ or
v$=\sqrt{1-4M_\chi ^2/s}=0$, the orbital angular momentum $L$ is
necessarily zero.  With two identical Majorana fermions, the
two-particle wave function must be antisymmetric under particle
interchange.  Since $L=0$ at threshold, the $\chi\chi$ spatial wave
function is even, and the wave function must be antisymmetrized in its
spin.  The antisymmetric spin wave function is the $S=0$ state.  Thus,
the only contributing partial wave at threshold is the $^1 S_0$ state.
We have just seen that this is also the only state with no v$^{2L}$
suppression, so one may expect an unsuppressed Majorana annihilation
rate at threshold if and only if there is a $^1 S_0$ partial wave.

One may also invoke $CP$ invariance to note that the spin $S$ of an
initial and final pair of spin-1/2 fermions, Dirac or Majorana, are the same.
This follows simply from $CP=(-)^{L+S} (-)^{L+1}=(-)^{S+1}$, and the
fact that $S=0,\,1$ are the only possibilities for a pair of spin 1/2
particles.

What does this mean for a two-Majorana initial state which annihilates to a two-fermion final state?
The implications are best recognized after a 
Fierz transformation of the two fermion bilinears to ``charge-retention''
order, i.e., to a $\chi$-bilinear and an f-bilinear.  
Among the basis fermion bilinears, the candidates are just the pseudo-scalar
$\Psibar i\gamma_5\Psi$~($0^{-+})$, the scalar
$\Psibar\Psi$~($0^{++}$), and the axial vector $\Psibar\gamma^\mu\gamma_5 \Psi$~($1^{++}$).  
The vector $\Psibar\gamma^\mu\Psi$ and tensor $\Psibar\sigma^{\mu\nu}\Psi$
bilinears are $C$-odd and therefore disallowed,
while the pseudo-tensor bilinear $\Psibar i\gamma_5\sigma^{\mu\nu}\Psi
=\frac{i}{2}\epsilon^{\mu]nu}_{\ \ \alpha\beta}\sigma^{\alpha\beta}$
does not couple to a Majorana current because $\sigma^{\alpha\beta}$ does not.
If the Fierz'd bilinears contain a pseudo-scalar, there is no suppression of the rate.  
Otherwise, there is a v$^2$ rate suppression.  
If the Fierz'd bilinears contain an axial vector piece, 
it offers a $(m_f/M_\chi)^2$-suppressed $s$-wave contribution,
unless accompanied by a radiated $W$ or $Z$ or $\gamma$.
The main results of this discussion of are summarized in Table~\rf{table:LSJ}.

\begin{table}
        \begin{tabular}{|c|c|c|c|c|c|c|c|c|}
        \hline\hline
        $\mathbf{L}$ & $\mathbf{S}$ & $\mathbf{P=(-)^{L+1}}$ & $\mathbf{C=(-)^{L+S}}$ & $\mathbf{^{2S+1}L_J}$  & $\mathbf{J^{PC}}$ & \textbf{Name } 
        		& \textbf{Dirac Op} & $\mathbf{v^{2L}}$ \\ \hline\hline
        \multicolumn{9}{|c|}{\textbf{C-even states}} \\ \hline
        0   & 0   & $-$     & +     & $^{1}S_0$     & $0^{-+}$    & pseudo-scalar     &   $i\gamma_5$                      &  v$^0$   \\ \hline
        1   & 1   & +         & +    & $^{3}P_0$     & $0^{++}$    &   scalar                  &   $1$                     	                  &  v$^2$   \\ \hline
        1   & 1   & +         & +    & $^{3}P_1$     & $1^{++}$    & axial-vector          & $\gamma_5\gamma^k$      &  v$^2$    \\ \hline
        0   & 0   & $-$	& +	 & $^{1}S_0$     & $0^{-+}$     & 		            & $\gamma_5\gamma^0$	&  v$^0$   \\ \hline
        \multicolumn{9}{|c|}{\textbf{C-odd states (unavailable for Majorana pair)}} \\ \hline\hline
        0   & 1   & $-$    & $-$   & $^{3}S_1$      & $1^{--}$    & vector              &  $\gamma^k$     &  v$^0$          \\  \hline
        1   & 0   & +        & $-$   & $^{1}P_1$      & $1^{+-}$   &  	   	       & $\gamma^0$      &  v$^2$          \\  \hline
        1   & 0   & +        & $-$   & $^{1}P_1$      & $1^{+-}$   & tensor              & $\sigma^{jk}$      & v$^2$   	 \\  \hline
        0   & 1   & $-$    & $-$   & $^{3}S_1$      & $1^{--}$    &   		       & $\sigma^{0k}$    &  v$^0$           \\  \hline\hline
        \multicolumn{9}{|c|}{\textbf{Redundant C-even states unavailable for Majorana pair}} \\ \hline
        0   & 0   & $-$     & +     & $^{1}S_0$    & $0^{-+}$     & pseudo-tensor     & $\gamma_5\sigma^{jk}$      &  v$^0$    \\ \hline
        1   & 1   & +         & +    & $^{3}P_1$     & $1^{++}$    & 			            & $\gamma_5\sigma^{0k}$     &  v$^2$    \\ \hline\hline 
        \end{tabular}
    \caption{{\bf Decomposition of fermion bilinear currents into $s$-channel partial waves.} 
    As we noted in the main text, the current piece $\gamma_5\gamma^0$ has a helicity-suppressed $s$-wave.
    We further note here that from the relation 
    $\gamma_5\sigma^{\mu\nu}=\frac{i}{2}\epsilon^{\mu\nu}_{\ \ \alpha\beta}\sigma^{\alpha\beta}$, one sees that 
    (i) the pseudo-tensor does not couple to Majorana fermions, and 
    (ii) one has 
    $P(\gamma_5\sigma^{\mu\nu})=P(\sigma^{\not\mu\not\nu})$, and 
    $C(\gamma_5\sigma^{\mu\nu})= - C(\sigma^{\not\mu\not\nu})$.
	    }
    \label{table:LSJ}
\end{table}

There is some subtlety associated with the s-channel exchange of an
axial-vector.  The axial-vector is an $L=1$ mode, and we have
seen that this mode elicits a v$^2$ suppression in the rate.  However,
the exchange particle is off-shell (away form resonance) and so has a
timelike pseudo-scalar piece in addition to the axial three-vector
piece which carries its polarization.  This pseudo-scalar coupling is effectively
$\partial_\mu\,(\Psibar\gamma^\mu\gamma_5\Psi)$.  The weak interaction
coupling of the pion to the axial vector current provides a familiar
example of such a coupling.  The axial current is not conserved, and
so the pseudo-scalar coupling is nonzero.  One has
$\partial_\mu\,(\Psibar\gamma^\mu\gamma_5\Psi)=2im_f\Psibar\gamma_5
\Psi -\frac{\alpha_W}{\pi}\epsilon^{\mu\nu\alpha\beta}k_\mu\lambda_\nu
(k) \kbar_\alpha\lambda_\beta (\kbar)$.  The first term shows an
$m_f$-dependence in the amplitude, leading to $(m_f/M_\chi)^2$
helicity-suppression of the $L=0$ piece, while the second term is the
famous anomalous VVA coupling.  It offers $W^+W^-$ and $ZZ$ production
(with momenta $k$, $\kbar$ and helicities $\lambda(k)$,
$\lambda(\kbar)$), but at higher order $\alpha_W= g_V^2/4\pi$ in the
electroweak $Wf\fbar$ or $Zf\fbar$ coupling $g_V$.  

It is illuminating to explain in this context the often seen remark
that LSP-annihilation has a helicity-suppressed rate to fermions.
This is true for SUSY extensions of the SM, but not true in general.
 For SUSY extensions of the SM, the
annihilation graphs consist of $t$-channel scalar exchanges, and from
crossing the identical Majorana fermions, also $u$-channel scalar
exchanges; in addition, there are scalar, pseudo-scalar and
axial-vector $s$-channel exchanges.  Fierzing the $t$- and $u$-channel
scalar exchanges yields $s$-channel axial-vector
bilinears, with the concomitant helicity-suppressed
$L=0$ contribution and v$^2$-suppressed $L=1$ contribution to the
annihilation rate.  The only contributions that are potentially large
come from the $s$-channel pseudo-scalars.  If the scalars and
pseudo-scalars are Higgs particles, then their Yukawa couplings $g_Y$ to the
SM fermion are all proportional to $(m_f/{\rm vev})$, thereby giving
the same effect as a true helicity suppression.  
A more general scalar or pseudo-scalar field, not
complicit in fermion mass generation, would couple with an arbitrary $g_Y$.
To summarize, with an
$s$-channel scalar or pseudo-scalar exchange, there is no
helicity suppression arising from the Yukawa couplings if they are  arbitrary.  
The pseudo-scalar exchange proceeds in the $L=0$ partial wave, with no v$^2$
suppression of the rate.  On the other hand, the scalar exchange
proceeds in the $L=1$ partial wave, which suppresses the
$\chi\chi$~annihilation rate by v$^2$.  
These deductions from partial wave analysis
hold true for annihilation of Dirac or Majorana DM.


\subsection{Projecting onto $s$-channel partial waves using Fierz Transformations}
\label{subsec:Fierz}
For fermionic dark matter, the Fierz transformations effect the natural projection of 
$2\rightarrow 2$ $t$- and $u$-exchange processes into partial waves. 
The $t$ and $u$~channel matrix elements for
annihilation, which are of the form $(\chibar \; \Gamma_A l)(\bar l \;
\Gamma_B \chi)$, are Fierzed into a sum of $s$-channel amplitudes of the ``charge-retention'' form
$(\chibar \; \Gamma_1 \chi)(\bar l \; \Gamma_2 l)$.
The Fierzed $s$-channel amplitudes are readily categorized into partial wave amplitudes,
and into fermion-pair spin states, by noting which Dirac operators $\Gamma_1$ and $\Gamma_2$ appear;
in turn, the spin states and the partial waves which appear determine whether the amplitudes are 
mass/helicity suppressed, velocity suppressed, or un-suppressed.  
In the rest of this section we develop these remarks.

\subsubsection{Fierz Transformations in the Chiral Basis}
\label{subsubsec:FierzIntro}
Helicity projection operators are essential in chiral gauge
theories, so it is worth considering the reformulation of the usual Fierz
transformations in the standard basis, to the chiral basis~\cite{Nishi:2004st}.  
(A discussion of Fierz transformations in the standard basis may be found in most textbooks on field theory;
details on the extension to the chiral basis presented here,
are available in  an appendix of~\cite{Bell:2010ei}.)
We place hats above the generalized Dirac matrices constituting the chiral basis.
The chiral basis is defined by the matrices
\beq{chiralbasis}
\{\Ghat^B\}=\{P_R,\,P_L,\,P_R\gamma^\mu,\,P_L\gamma^\mu,\,\half\sigma^{\mu\nu} \}\,,
\quad {\rm and} \quad 
\{\Ghat_B\}=\{P_R,\,P_L,\,P_L\gamma_\mu,\,P_R\gamma_\mu,\,\half\,\sigma_{\mu\nu} \}\,,
\eeq
where $P_R\equiv\half (1+\gamma_5)$ and $P_L\equiv\half (1-\gamma_5)$ 
are the usual helicity projectors.
Notice that the dual of $P_R\gamma^\mu$ 
is $P_L\gamma_\mu$, and the dual of $P_L\gamma^\mu$ is $P_R\gamma_\mu$.
The tensor matrices in this basis contain factors of $\half$:
$\Ghat^T=\half\sigma^{\mu\nu}$ and $\Ghat_T = \half\sigma_{\mu\nu}$.
These facts result from the orthogonality and normalization properties
of the chiral basis and its dual.
In the chiral basis, one finds a completeness relation
%
$(\openone)\,[\openone] = \half\,(\Ghat_B\,]\,[\Ghat^B\,) = \half\,(\Ghat^B\,]\,[\Ghat_B\,)$
which effects a master formula: the outer product of {\sl any} two  $4\times 4$ matrices 
$\X$ and $\Y$ can be expressed in terms of the   Fierzed forms of the 
chiral basis matrices:
\beq{chiralXandY}
(\X)\,[\Y] = (\X\openone)\,[\openone \Y]=\half (\X\,\Ghat_B\,\Y\,]\,[\,\Ghat^B\,) 
   = \quarter\,Tr\,[\X\,\Ghat^C\,\Y\,\Ghat_B ]\ (\Ghat^B\,]\,[\Ghat_C\,)\,.
\eeq
The parentheses symbols here are a convenient shorthand~\cite{Taka1986} for matrix
indices, in an obvious way.  
Relevant to $2\rarr 2$ fermion scattering is the 
substitution $\X=\Ghat^D$ and $\Y=\Ghat_E$ into Eq.~\rf{chiralXandY}. 
One gets
\beq{chiralFierz1}
(\Ghat^D )\,[\Ghat_E ]=\frac{1}{4}\,Tr\,[\Ghat^D\,\Ghat^C\,\Ghat_E\,\Ghat_B ]\ (\Ghat^B ]\;[\Ghat_C )\,,
\eeq
and evaluating the trace in Eq.~\rf{chiralFierz1} leads to the following $2\rarr 2$ Fierz transformation
matrix in the chiral-basis:
\bea{chiralFierz}
\left(
\barr{c}
(P_R)\ [P_R] \\
(P_L)\ [P_L] \\
(\hT)\ [\hT] \\
\!\!(\gamma_5\,\hT)\ [\hT] \\
(P_R)\ [P_L] \\
(P_R\gamma^\mu)\ [P_L\gamma_\mu] \\
(P_L)\ [P_R] \\
(P_L\gamma^\mu)\ [P_R\gamma_\mu] \\
(P_R\gamma^\mu)\ [P_R\gamma_\mu] \\
(P_L\gamma^\mu)\ [P_L\gamma_\mu] \\
\earr
\right) 
= \quarter
\left(
\barr{rrrr|rr|rr|r|r}
 2  & 0 & 1 & 1 &   &   &   &   &   &   \\ 
 0  & 2 & 1 &-1 &   &   &   &   &   &   \\
 6  & 6 &-2 & 0 &   &   &   &   &   &   \\
 6  &-6 & 0 & 2 &   &   &   &   &   &   \\ \hline
    &   &   &   & 0 & 2 &   &   &   &   \\
    &   &   &   & 8 & 0 &   &   &   &   \\ \hline
    &   &   &   &   &   & 0 & 2 &   &   \\
    &   &   &   &   &   & 8 & 0 &   &   \\ \hline
    &   &   &   &   &   &   &   &-4 & 0 \\ \hline
    &   &   &   &   &   &   &   & 0 &-4    
\earr
\right)
\ \left(
\barr{c}
(P_R]\ [P_R) \\
(P_L]\ [P_L) \\
(\hT]\ [\hT) \\
\!\!(\gamma_5\,\hT]\ [\hT) \\
(P_R]\ [P_L) \\
(P_R\gamma^\mu]\ [P_L\gamma_\mu) \\
(P_L]\ [P_R) \\
(P_L\gamma^\mu]\ [P_R\gamma_\mu) \\
(P_R\gamma^\mu]\ [P_R\gamma_\mu) \\
(P_L\gamma^\mu]\ [P_L\gamma_\mu) \\
\earr
\right)\,.
\eea
Non-explicit matrix elements in Eq.~\rf{chiralFierz} are zero,
and we have introduced a shorthand $\hT$ for 
either $\Ghat^T=\half\sigma^{\mu\nu}$ or  $\Ghat_T=\half\sigma_{\mu\nu}$.
In Eq.~\rf{chiralFierz} we have included one non-member of the basis set,
namely $\gamma_5\,\hT$; it is connected to $\hT$ via the relation  
%
$\gamma_5\,\sigma^{\mu\nu} = \frac{i}{2}\epsilon^{\mu\nu\alpha\beta}\sigma_{\alpha\beta}$.
Explicit use of $\gamma_5\,\hT$ in Eq.~\rf{chiralFierz} is an efficient way to express
the chiral Fierz transformation.
The block-diagonal structures, delineated with horizontal and vertical
lines, show that ``mixing'' occurs only within the subsets 
$\{ P_R\otimes P_R,\ P_L\otimes P_L,\ \hT\otimes \hT,\ \gamma_5\,\hT\otimes
\hT\}$, and $\{P_R\otimes P_L,\ P_R\gamma^\mu\otimes P_L\gamma_\mu \}$.
The importance of the chiral transformation Eq.~\rf{chiralFierz} for us is that it converts
$t$-channel and $u$-channel exchange graphs into $s$-channel form, for
which it is straightforward to evaluate the partial waves.  

We note that Eq.~\rf{chiralXandY} lends itself to a generalization of the 
Fierz transformation beyond the $2\rarr 2$ case.
One has only to let $\X$ and $\Y$ represent fermion lines with radiative emissions. 
Such a generalization, offering insight into the non-suppressed $2\rarr 3$ amplitude~\cite{Bell:2010ei},
is presented in the next section of this talk.

\subsubsection{Majorana Dark Matter}
\label{subsubsec:Majorana DM}
Now I specialize the discussion of Fierz transformations to the case of Majorana dark matter.
A Majorana particle is invariant under charge conjugation ${\cal C}$.
Accordingly, the Majorana field creates and annihilates the same particle.
This implies that for each $t$-channel diagram (shown in Fig.~\ref{fig:feyngraphs_ace}),
there is an accompanying $u$-channel diagram (shown in Fig.~\ref{fig:feyngraphs_bdf}).
The latter is obtained by interchanging the momentum and spin of the two Majorana fermions.
The relative sign between the $t$- and $u$-channel amplitudes is $-1$ in accord with Fermi statistics.
As is well known, this relative minus sign has important consequences for Majorana DM.
For example,
consider the Fierzed (i.e., $s$-channel) bilinear for $\chi$-annihilation:
$\bar{v}(k_1,s_1) \Gamma_B u(k_2,s_2)$.
The associated Fierzed bilinear from the $(k_1\leftrightarrow k_2)$-exchange graph, 
with its relative minus sign, is 
$-\bar{v}(k_2,s_2) \Gamma_B u(k_1,s_1)$.
Constraints relate the four-component Dirac
spinors to their underlying two-component Majorana spinors.  
These constraints, any one of which implies the other three, are
\beq{Majspinors}
 u(p,s) = C\bar{v}^T(p,s)\,,\quad  \bar{u}(p,s)=-v^T(p,s)C^{-1}\,,\quad
 v(p,s) = C\bar{u}^T(p,s)\,,\quad  \bar{v}(p,s) = -u^T(p,s)C^{-1}\,,
\eeq
where $C$ is the charge conjugation matrix. 
These Majorana conditions on the spinors allow us to rewrite the exchange bilinear as 
(suppressing spin labels for brevity of notation)
%
\be
\label{Majswap}
-\bar{v}(k_2)\Gamma_B u(k_1) = u^T(k_2)C^{-1}\Gamma_B C\bar{v}^T(k_1)
= \left[ \bar{v}(k_1)(C^{-1}\Gamma_B C)^T u(k_2)\right]^T
= \bar{v}(k_1) (\eta_B\Gamma_B) u(k_2)\,.
\ee
%
For the final equality, we have used (i) the fact that the transpose symbol can be dropped from a number,
and (ii) the identity $(C^{-1}\Gamma_B C)^T=(\eta_B (\Gamma_B)^T)^T =\eta_B\Gamma_B$,
where  $\eta_B=+1$ for $\Gamma =$ scalar, pseudo-scalar, axial vector, and 
$\eta_B = -1$ for $\Gamma =$ vector or tensor.
In the four-Fermi or heavy propagator limit, where the differing
momenta in the $t$- and $u$-channel propagators can be ignored, one
obtains an elegant simplification.  
Subtracting the u-channel amplitude from the t-channel amplitude, one
arrives at the weighting factor $(1+\eta_B)$, which is two for $S,P,$
and $A$ couplings, and zero for $V$ and $T$ couplings.  
This means that $V$ and $T$ couplings in the Fierzed bilinears of the $\chi$-current are 
identically zero in the four-Fermi limit.
Including the differing $t$ and $u$ channel propagators, 
one gets a nonzero residual proportional to 
$(t-M_\eta^2)^{-1} - (u-M_\eta^2)^{-1}= s{\rm v}\cos{\bar\theta}/(t-M_\eta^2) (u-M_\eta^2)$,
where ${\bar\theta}$ is the $2\rarr 2$ scattering angle in the CoM.
This residual is higher order in~v, and so generally negligible.
Thus, after Fierzing, only the axial vector coupling of the
$\chi$-current is significant, and the factor of $1+\eta_A=2$ multiplies
the (7-8)-element~(one-half) in the Fierz matrix of
Eq.~\rf{chiralFierz} to give a net weight of~1.

One sees from Eq.~\rf{chiralFierz} that the bilinear current $2(P_L) [P_R]$ Fierzes to $8(P_L\gamma^\mu] [P_R\gamma_\mu)$.
Since we drop the vector current for Majorana particles, we are left with pure axial-vector couplings. Next we show 
how a pure axial-vector current in the $s$-wave amplitude implies a helicity-suppression.

\subsection{Class of Models for which $\chi\chi\rarr\ell\ellbar$ Annihilation is Suppressed }
\label{subsec:suppression}

\begin{table} 
\begin{tabular}{||c|c||c|c||c|c||}
\hline\hline
\multicolumn{2}{||c||}{s-channel bilinear $\Psibar\,\Gamma_D\,\Psi$} 
	& \multicolumn{2}{c||}{${\rm v}=0$ limit (projects out pure $s$-wave)} 
   	& \multicolumn{2}{c||}{${\rm M}=0$ limit (zeroes out helicity suppression)} \\ \cline{3-6}
\multicolumn{2}{||c||}{} & parallel spinors & antiparallel spinors & parallel spinors & antiparallel spinors \\  \hline\hline
scalar        & $\Psibar\,\Psi$                         
   & 0 & 0 & $ \sqrt{s}$ & 0 \\ \hline\hline
pseudo-scalar & $\Psibar\,i\gamma_5\,\Psi$               
   & $-2iM$ & 0 & $-i\sqrt{s}$ & 0   \\ \hline\hline
axial-vector  & $\Psibar\,\gamma_5\,\gamma^0\,\Psi$     
   & $2M$ & 0 & 0 & 0   \\ \cline{2-6}
              & $\Psibar\,\gamma_5\,\gamma^j\,\Psi$                                     
   & 0 & 0 & 0 & $\sqrt{s}\,(\pm\delta_{j1}-i\delta_{j2})$ \\ \hline\hline
vector        & $\Psibar\,\gamma^0\,\Psi$               
   & 0 & 0 & 0 & 0 \\ \cline{2-6}
              & $\Psibar\,\gamma^j\,\Psi$          
   & $\mp 2M\,\delta_{j3}$ & $-2M\,(\delta_{j1}\mp i\delta_{j2})$ & 0 & $-\sqrt{s}\,(\delta_{j1}\mp i
\delta_{j2})$ \\ \hline\hline
tensor        & $\Psibar\,\sigma^{0j}\,\Psi$        
   & $\mp 2iM\,\delta_{j3}$ & $-2iM\,(\delta_{j1}\pm\delta_{j2})$ & $-i\sqrt{s}\,\delta_{j3}$ & 0 \\ \cline
{2-6}
        & $\Psibar\,\sigma^{jk}\,\Psi$        
   & 0 & 0 & $\pm\sqrt{s}\,\delta_{j1}\delta_{k2}$ & 0 \\ \hline\hline
pseudo-tensor  & $\Psibar\,\gamma_5\,\sigma^{0j}\,\Psi$ 
   & 0 & 0 & $\pm i\sqrt{s}\,\delta_{j3}$ & 0 \\ \cline{2-6}
   & $\Psibar\,\gamma_5\,\sigma^{jk}\,\Psi$ 
   & $\mp 2M\,\delta_{j1}\delta_{k2}$ & $-2M\,(\delta_{j2}\delta_{k3}\mp i\delta_{j3}\delta_{k1})$ 
         & $-\sqrt{s}\,\delta_{j1}\delta_{k2}$ & 0 \\ \hline\hline
\end{tabular}
\caption{
Extreme non-relativistic and extreme relativistic limits for s-channel fermion bilinears.
In order for a term with an initial-state DM bilinear and a final-state SM bilinear to remain unsuppressed,
the DM bilinear must have a nonzero entry in the appropriate cell of the ``${\rm v}=0$ limit'' columns,
and the SM bilinear must have a non-zero term in the appropriate cell of the  ``${\rm M}=0$ limit'' columns. 
Otherwise, the term is suppressed. 
We recall that antiparallel spinors correspond to parallel particle spins 
(and antiparallel particle helicities for the $M=0$ current), and vice versa.
Amplitudes are shown for $\ubar\,\Gamma_D\,v = [ \vbar\,\Gamma_D\,u ]^*$.
The two-fold $\pm$ ambiguities reflect the two-fold spin assignments for parallel spins, and 
separately, for antiparallel spins.
}
\label{table:bilinearlimits}
\end{table}

\noindent
Consider products of s-channel bilinears in Fierz-transformed, charge-retention order 
$(\chibar \;\Gamma_1 \chi)(\bar \ell \; \Gamma_2 \ell)$.
To further address the question of which products of currents are
suppressed and which are not, we set ${\rm v}^2$ to zero in the
$\chi$-current, and $m_\ell^2$ to zero in the lepton current, and ask
whether the product of these currents is suppressed.  If the product of
currents is non-zero in this limit, the corresponding amplitude is
unsuppressed.  In Table~\ref{table:bilinearlimits} we give the results
for the product of all standard Dirac bilinears.  
Suppressed bilinears enter this table as zeroes.
(The derivation of these results is outlined in another appendix of~\cite{Bell:2010ei}.)

One can read across rows of this table to discover that the only
unsuppressed $s$-channel products of bilinears for the $2\rarr 2$
process are those of the pseudo-scalar, vector, and tensor.  (For
completeness, we also show results for the pseudo-tensor bilinears,
although the pseudo-tensor is not independent of the tensor, as delineated below
Eq.~\rf{chiralFierz}.)  For Majorana DM, the vector and
tensor bilinears are disallowed by charge-conjugation arguments 
and one is left with just the unsuppressed pseudo-scalar.

Let us now consider $t$-channel or $u$-channel processes.
Any $t$-channel or $u$-channel diagram that Fierz's to an $s$-channel
form containing a pseudo-scalar coupling will have an unsuppressed
$s$-wave amplitude.  From the matrix in Eq.~(\ref{chiralFierz}),
one deduces that such will be the case for any $t$- or $u$-channel
current product on the left side which finds a contribution in the
$1^{st}$, $2^{nd}$, $5^{th}$, or $7^{th}$ columns of the right side.
This constitutes the $t$- or $u$-channel tensor, same-chirality
scalar, and opposite chirality vector products (rows 1 through 4, and
6 and 8 on the left).
On the other hand, the $t$- or $u$-channel opposite chirality scalars
or same-chirality vectors (rows 5, 7, 9, and 10 on the left) do {\it 
not} contain a pseudo-scalar coupling after Fierzing to $s$-channel form.  
Rather, it is the suppressed axial-vector and 
vector (the latter for Dirac fermions only) that appear.

Interestingly, a class of the most popular models for fermionic DM 
annihilation to charged leptons or to quarks falls into this latter,
suppressed, category. It is precisely the opposite-chirality $t$- or
$u$-channel scalar exchange that appears in these models, an explicit
example of which will be discussed below.
Thus it is rows 5 and 7 in Eq.~(\ref{chiralFierz}) that categorize the common
model we will analyze.  After Fierzing to $s$-channel form, it is seen
that the Dirac bilinears are opposite-chirality vectors 
(i.e., linear combinations of $V$ or $A$).  
Dropping the vector term from the Majorana $\chi$-current we see that
the $2\rightarrow 2$ process couples a $C$-even $L=1$ axial vector 
$\chi$-current to a relativistic SM fermion-current which is an equal mixture
of $A$ and~$V$.
Reference to Table~\rf{table:bilinearlimits} reveals that the only component of the axial current 
which is non-vanishing in the $s$-wave (${\rm v}=0$ limit) is ${\bar\chi}\gamma_5\gamma^0\chi$.
However, there is no corresponding non-vanishing current $\bar\Psi\gamma_5\gamma^0\Psi$ or 
$\bar\Psi\gamma^0\Psi$ in the Table.  
Thus, the $s$-wave amplitude must be helicity suppressed.
Table~\rf{table:bilinearlimits} also reveals that a coupling of ${\bar\chi}\gamma_5\gamma^0\chi$ 
to $\bar\Psi\gamma_5\gamma^j\Psi$ or $\bar\Psi\gamma^j\Psi$ is helicity un-suppressed,
but requires a spin flip from parallel spinors in the initial state to antiparallel spinors in the final state.
Such a direct coupling would violate Lorentz invariance.
To the rescue comes  gauge boson emission,
which alters the fermion-pair spin state by one unit of helicity,
and couples ${\bar\chi}\gamma_5\gamma^0\chi$ 
to $\bar\Psi\gamma_5\gamma^j\Psi$ and $\bar\Psi\gamma^j\Psi$,
as we shall see.
An un-suppressed $s$-wave amplitude will be the result. 
%

We now explain why this $t$- or $u$-channel scalar exchange
with opposite fermion chiralities at the vertices is so common.  It
follows from a single popular assumption, namely that the dark matter
is a gauge-singlet Majorana fermion.  As a consequence of this
assumption, annihilation to SM fermions, which are $SU(2)$ doublets or singlets,
requires either an $s$-channel singlet boson, or a $t$- or $u$-channel
singlet or doublet scalar that couples to $\chi$-$f$.  In the first
instance, there is no symmetry to forbid a new force between SM
fermions, a disfavored possibility.  In the second instance, unitarity
fixes the second vertex as the hermitian adjoint of the first.  Since
the fermions of the SM are left-chiral doublets and right-chiral
singlets, one gets chiral-opposites for the two vertices of the $t$-
or $u$-channel.

Supersymmetry provides an analog of such a model.  In this case the
dark matter consists of 
Majorana neutralinos, which
annihilate to SM fermions via the exchange of (``right''- and
``left''-handed) $SU(2)$-doublet slepton and squark fields.  
In fact, the 1983 implementation of DM supersymmetric photinos
provided the first explicit calculation of $s$-wave suppressed Majorana annihilation~\cite{Haim1983}.
However, the class of models described above is more general than just supersymmetric models.

We may illustrate our arguments with a minimal leptophilic interaction in which 
the DM consists of gauge-singlet Majorana fermions $\chi$ which
annihilate to SM leptons via the $SU(2)$-doublet mediator fields $\eta$
\begin{equation}
g\left(\nu\,\ell^-\right)_L\,\left(\varepsilon\right)\,
\left(
\barr{l}
\eta^+ \\
\eta^0 \\
\earr
\right)\chi + h.c.
= g(\nu_L\eta^0 - \ell_L \eta^+)\chi + h.c.
\label{eq:ma}
\end{equation}
Here $g$ is a coupling constant, $\varepsilon$ is the
$2\times 2$ antisymmetric matrix which projects the SU(2) singlet out from the addition of two doublets, 
and $(\eta^+$, $\eta^0)$ form the $SU(2)$ doublet scalar which mediates the annihilation. 
As discussed above, the $u$- and $t$-channel amplitudes for DM
annihilation to leptons, of the form $(\chibar P_L \ell)\,(\bar\ell P_R\chi)$, 
become pure $(\chibar P_L\gamma^\mu \chi)\,(\bar\ell P_R\gamma_\mu \ell)$ 
under the chiral Fierz transformation.
The product of the Majorana and fermion bilinears then leads to an
$AA$ term and an $AV$ term. However, reference to
Table~\ref{table:bilinearlimits} shows that neither of these terms
leads to an unsuppressed amplitude: in all cases, either the lepton
bilinear is suppressed by $m_\ell$ in the $s$-wave or the DM bilinear by v in the $p$-wave.
Thus, Majorana DM annihilation to a lepton pair is suppressed in this class of model.

\section{Lifting the Suppression}
\label{sec:unsuppression}


Allowing the lepton bilinear to radiate a $W$ or $Z$ boson (as shown in Figs.~\rf{fig:feyngraphs_ace} and~\rf{fig:feyngraphs_bdf}) 
does yield an unsuppressed amplitude.
The inevitable question is ``Why?"
Physically, the un-suppression works because the gauge boson carries
away a unit of angular momentum,
allowing a fermion spin-flip such that there is no longer a mismatch between
the chirality of the leptons and their allowed two-particle spin state.
But this is a glib answer, for we know that emission of a gauge boson in the four-Fermi limit (i.e. FSR but no IB) 
does not un-suppress the $s$-wave amplitude.
A deeper answer is warranted.

\subsection{Generalized Fierz  Transformation}
\label{generalizedFierz}
To address this question mathematically, 
we invoke the more general Fierz rearrangement applicable to $2\rarr 2\,{\rm fermion\ }+(N-2)\,{\rm bosons}$ processes.  
In the helicity basis, this generalized Fierz equation is given by Eq.~\rf{chiralXandY}.
The analog equation in the standard (non-chiral) basis is~\cite{Bell:2010ei}
\beq{genFierz}
(\X)\,[\Y] =  \frac{1}{4^2}\,Tr\,[\X\,\Gamma^B\,\Y\,\Gamma_C ]\ (\Gamma^C\,]\ [\Gamma_B\,)\,,
\eeq
where $\X$ and $\Y$ are any $4\times 4$ matrices.
From Table~\rf{table:bilinearlimits} we see that setting $\Gamma^C$ to $\gamma_5\,\gamma^0$,
the only structure available to a non-relativistic Majorana current other than the pseudo-scalar,
and setting $\Gamma_B$ to either $\gamma^j$ or $\gamma_5\gamma^j$, 
provides an unsuppressed product of the Majorana DM bilinear and the 
charged lepton bilinear.
Moreover, for the $W/Z$-bremsstrahlung process, we input the un-Fierzed couplings 
$P_L$ and $q^{-2}\,P_R\,\slashed{q}\,P_L\,\slashed{\epsilon}$ for $X$ and $Y$, respectively,
where $\epsilon$ is the gauge boson polarization vector.
So a necessary condition for the radiative process to be un-suppressed is that 
$Tr\,[P_L\,\gamma^j\,P_R\,\slashed{q}\,P_L\,\slashed{\epsilon}\,\gamma_5\,\gamma_0 ]$ is unsuppressed. 
This trace reduces to  $Tr\,[P_R\,\gamma_0\,\gamma^j\,\slashed{q}\,\slashed{\epsilon} ]$.
We note that the coupling of a $\gamma^0$-current to a $\gamma^j$-current is just the recipe 
suggested for $s$-wave un-suppression in the preceding section.
The expansion of this trace as scalar products contains terms such as $q_0\epsilon^j$
and $({\vec\epsilon}\times{\vec q})^j$, which are nonzero and unsuppressed by fermion masses.
Thus, the $2\rarr 3$~process contains an unsuppressed $s$-wave amplitude.
Not apparent in this argument is the need for IB.  
To establish that, we must calculate this unsuppressed amplitude directly,
which we did~\cite{Bell:2011if}.


\subsection{The Full Calculation of s-wave {\boldmath $W/Z$}-bremsstrahlung}
\label{subsec:Wbrehms}
To demonstrate the un-suppression of the $s$-wave amplitude, 
it is sufficient and convenient to take the limits $m_\ell = 0$ (hereby removing any option of helicity suppression) and 
${\rm v}=0$ (projecting out just the $s$-wave).
In our calculation, we invoke both limits, and for simplicity we set  the mediator masses $m_{\eta^\pm}$ and $m_{\eta^0}$ to be equal.
In the limit of vanishing lepton masses, 
the Ward identity for EW~bremsstrahlung takes the same form as for photon bremsstrahlung 
since the EW axial-vector current is conserved in this limit of massless fermions.  
Thus, defining the EW~bremsstrahlung amplitude by
$\mathcal{M} =\mathcal{M}_{\rm EW}^\mu\epsilon_\mu(Q)$,
the Ward identity requires that
$Q_\mu \mathcal{M}_{\rm EW}^\mu =0$. 
We have checked that this form of the Ward identity is separately satisfied for the diagram sets (a)+(c)+(e) and (b)+(d)+(f),
thus demonstrating the independent gauge invariance of each subset.

Note that $t$- (Fig.~\ref{fig:feyngraphs_ace}) and $u$-channel (Fig.~\ref{fig:feyngraphs_bdf}) amplitudes of the six contributing Feynman diagrams  
are simply related by the $k_1\leftrightarrow k_2$ interchange symmetry.
The full amplitude is the sum of the partial amplitudes, properly weighted
by a minus sign when two fermions are interchanged. Thus we have
$\mathcal{M}_{\rm EW} = (\mathcal{M}_a + \mathcal{M}_c + \mathcal{M}_e)
- (\mathcal{M}_b + \mathcal{M}_d + \mathcal{M}_f)
= (\mathcal{M}_a + \mathcal{M}_c + \mathcal{M}_e) - (k_1\leftrightarrow k_2)$.
Diagrams (e) and (f) explicitly demonstrate the IB.

There is a subtlety worthy of explanation.
The IB graphs split the internal propagator into two propagators,
reducing the amplitude by an ``extra'' power of $M_\chi^2/M_\eta^2$.
This suggests that the leading order contribution to EW~bremsstrahlung comes from the four graphs (a)-(d),
as would result in the four-Fermi limit of the scattering 
(wherein the internal propagator is ``pinched'' to a single spacetime point).  
Although this notion is alluring, it is wrong.  
As revealed in~\cite{Ciafaloni:2011sa} and \cite{Bell:2011if}, the Ward identity for these four graphs 
leads to a vanishing bremsstrahlung amplitude.
Thus, the IB graphs are essential, and the four-Fermi limit of the DM-SM fermion interaction fails to include 
$W/Z$ or $\gamma$~bremsstrahlung.

We also find that the longitudinal polarization of the $W$ does not contribute to 
the $s$-wave amplitude, i.e.
$\mathcal{M}^\mu_{\rm EW} \epsilon_{L\,\mu}(Q) = 0$\,,
in our limit of degenerate mediator masses ($M_{\eta^\pm}=M_{\eta^0}$) in the IB diagrams.
Therefore, in this mass-degenerate limit the $W$ boson behaves as a massive transverse photon
with just two transverse polarization states, and our calculation of 
$W$~bremsstrahlung must reduce to the known results for photon bremsstrahlung in the 
$m_W\rightarrow 0$ limit, modulo coupling constants.  
We have checked that this happens.

%
\begin{figure}[t]
\includegraphics[height=0.20\textheight,width=0.32\columnwidth]{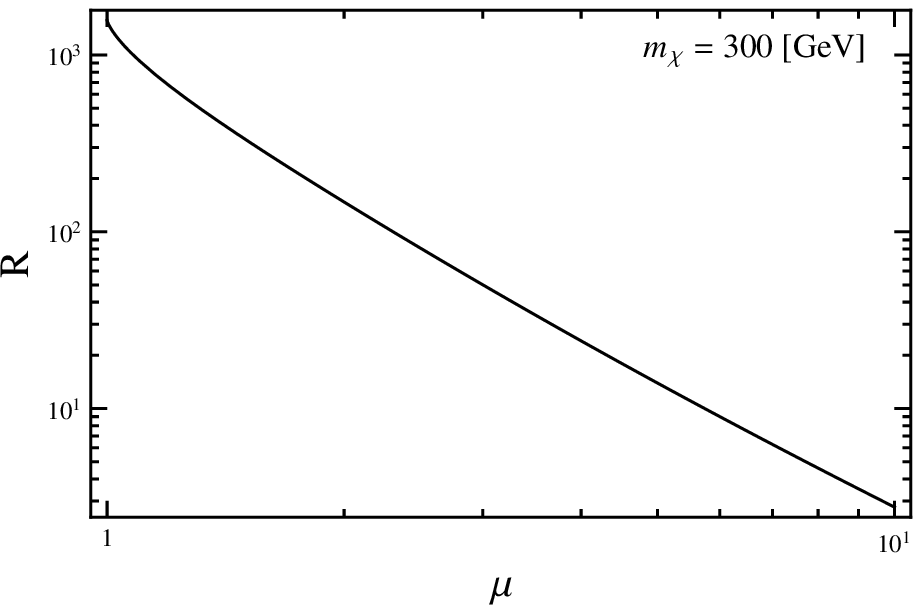}
\includegraphics[height=0.20\textheight,width=0.32\columnwidth]{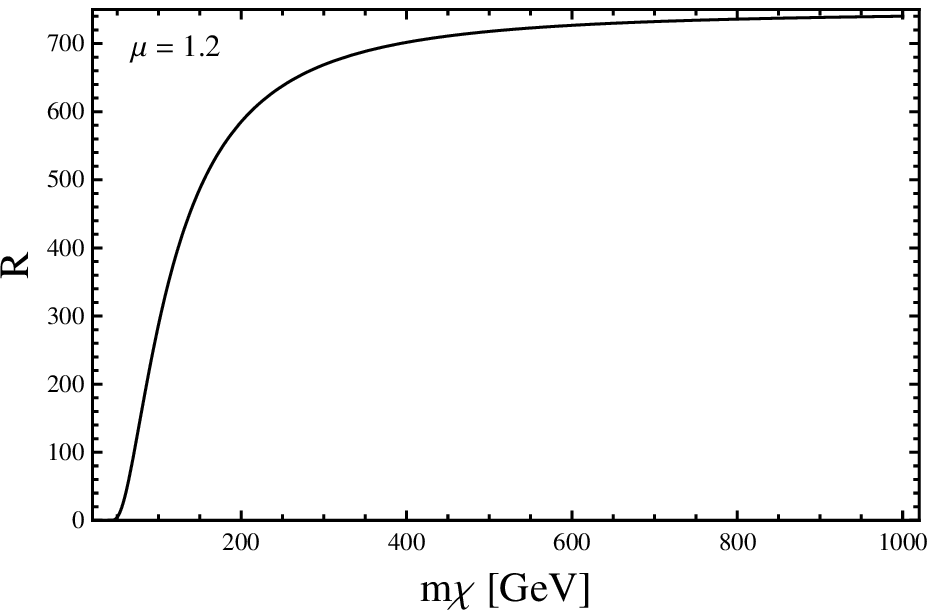}
\includegraphics[height=0.20\textheight,width=0.32\columnwidth]{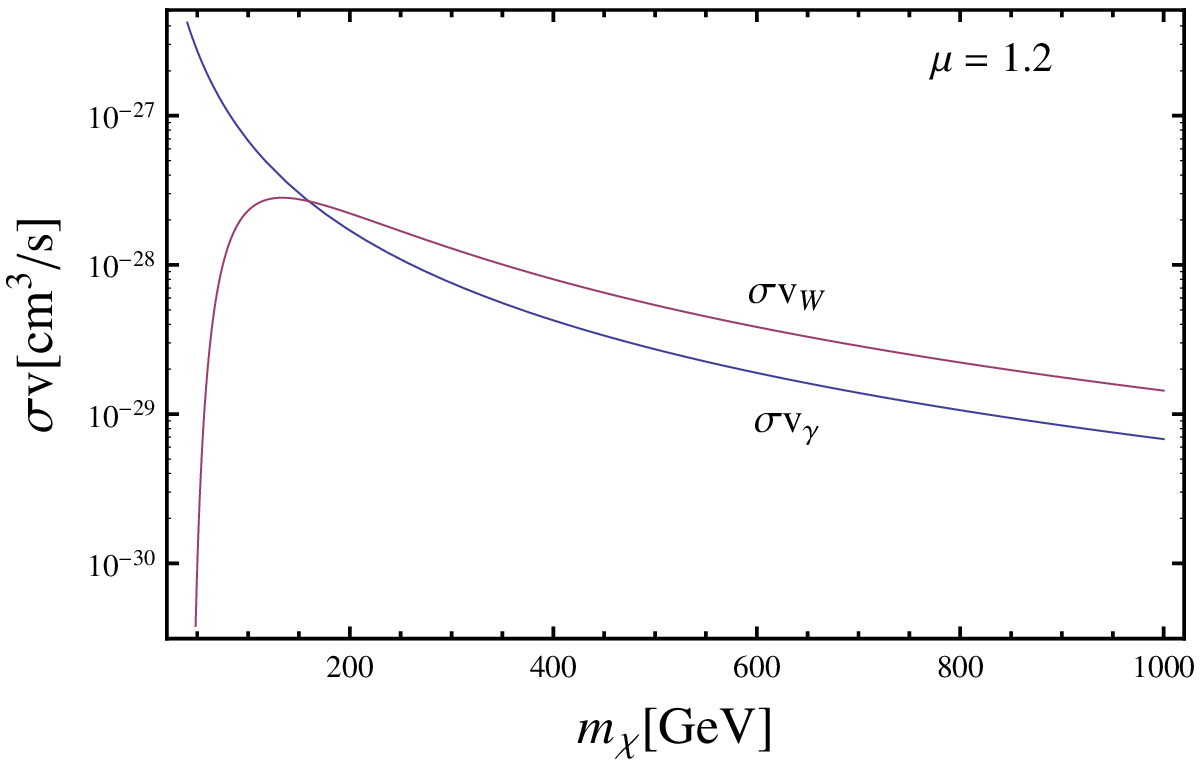}
\caption{
\label{fig:rates} 
The ratio $R={\rm v}\,\sigma (\chi\chi\to e^+ \nu W^-) /{\rm v}\,\sigma (\chi\chi\to e^+e^-)$ 
versus $\mu = (M_\eta/M_\chi)^2$ (left) and $M_\chi$ (center), for
${\rm v} = 10^{-3}c$, and fixed $M_\chi=300$~GeV and $\mu=1.2$~GeV, respectively.
The rate for $\chi\chi\to e^+ \nu W^-$ and $\to e^+ e^- \gamma$ (right), 
for $\mu=1.2$ and unit $\chi\chi\eta$~coupling; 
for large DM mass, the rates differ by a factor of $1/(2
\sin^2\theta_W)=2.17$.
}
\end{figure} 

%

We find that the EW~bremsstrahlung rate falls monotonically with
increasing $M_\eta$ (or $\mu\equiv(M_\eta/M_\chi)^2$). 
This monotonic fall is shown in
Fig.~\ref{fig:rates}, where we plot the ratio of the $W$-strahlung rate to that of the lower order $2\rarr 2$ process, 
$R = \langle{\rm v}\,\sigma(\chi\chi\to e^+ \nu W^-)\rangle /\langle{\rm v}\,\sigma (\chi\chi\to e^+e^-)\rangle$.  The
lowest order process itself falls as $~ \mu^{-2}$, so the
$W$-strahlung process falls as $~\mu^{-4}$.  This latter dependence is
expected for processes with two propagators each off-shell by
$1/\mu$, thereby signaling leading order cancellations among FSR sub-diagrams (a)-(d) of Figs.~\rf{fig:feyngraphs_ace} and~\rf{fig:feyngraphs_bdf}.
Importantly, the effectiveness of the $W$-strahlung processes in
lifting suppression of the annihilation rate is evident in Fig.~\ref{fig:rates}.
In the first panel we see that the although the ratio is maximized for $\mu$ close to 1
where $M_\chi$ and $M_\eta$ are nearly degenerate,
the $W$-strahlung process still dominates over the
tree level annihilation even if a mild hierarchy between $M_\chi$ and
$M_\eta$ is assumed.  
The ratio exceeds 100 for $\mu \alt 2$, and exceeds unity for $\mu \alt 6$.
The second panel reveals that the ratio $R$ is
insensitive to the DM mass, except for low $M_\chi$ where the $W$ mass
significantly impacts phase space; one gleans that 
for $M_\chi\agt 3\,m_W$, the ratio $R$ is already near to its asymptotic value.

Presented in the right panel of the Fig.~\ref{fig:rates}
is a comparison of the $W$-strahlung rate to that for photon bremsstrahlung.  
For high dark matter masses where the $W$ mass is negligible, 
the two rates are identical except for the overall normalization, which is higher for $W$-strahlung 
by the factor $1/(2\sin^2\theta_W)=2.17$ .  
Another factor of two is gained for $W$-strahlung when the $W^+$ mode is added 
to the $W^-$ mode shown in the figure.

As remarked in the introduction, the correct dark matter energy fraction is obtained for
early-Universe thermal decoupling with an annihilation cross section
of $3\times 10^{-26} \textrm{cm}^3$/s.  It is seen in the right panel of Fig.~\ref{fig:rates} 
that the rate for the $s$-wave $W$-strahlung mode lies 2-3 orders of magnitude below this value.
Hence, radiative $W$-strahlung with its natural suppression factor $\alpha_W/4\pi$ is
probably not the dominant annihilation mode responsible for
early-Universe decoupling of Majorana dark matter.

A very interesting feature of $W$-bremsstrahlung emerges when the two intermediate scalars 
in the IB graph are not degenerate in mass.
The authors of Ref.~\cite{Garny:2011cj} found that with non-degenerate mediator masses, 
the longitudinal $W_L$  is in fact radiated. 
Moreover, depending on the size of the mass-squared splitting $M_{\eta^\pm}^2 -M_{\eta^0}^2$,
the rate of longitudinal~$W_L$ radiation may even exceed the radiation rate of the transverse~$W_T$.


\section{Importance of {\boldmath$W/Z$}-bremsstrahlung for Cosmic Signatures}
\label{sec:signatures}
\begin{figure*}[ht]
\includegraphics[height=0.26\textheight,width=0.6\columnwidth]{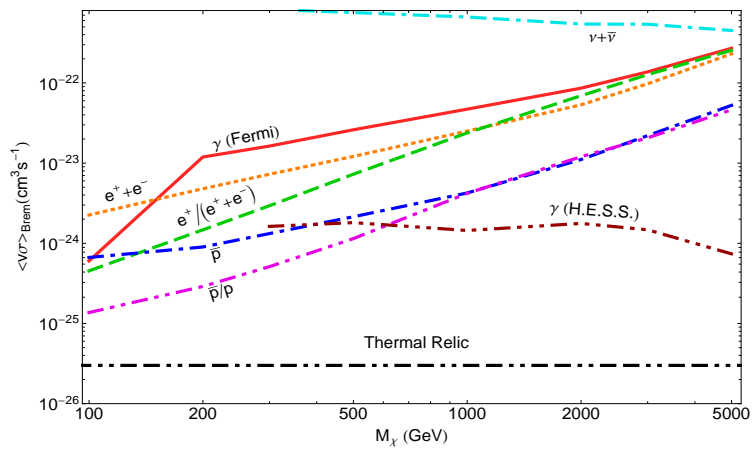}
\caption{Upper limits on $\langle v \sigma \rangle_{\rm Brem}$ using
  the `med' diffusion parameter set.  Shown are constraints based on
  the Fermi extragalactic background light (solid, red), 
  $e^++e^-$ flux (dots, orange),
  $e^+/(e^++e^-)$ ratio (dashes, green), 
  $\bar p$ flux (dot-dashes, blue), 
  $\bar p/p$ ratio (dot-dot-dashes, magenta),
  H.E.S.S. gamma rays (dot-dot-dot-dashes, maroon),
  and neutrinos (dot-dash-dashes, cyan).    
  Also shown for comparison is the
  expected cross section for thermal relic dark matter, $3\times
  10^{-26} \,\rm{cm}^3$/s. 
\label{fig:limits-brem}}
\end{figure*}

In Ref.~\cite{Bell:2011eu}, we presented the spectra of
stable annihilation products produced via $\gamma/W$/$Z$-bremsstrahlung.
After modifying the fluxes to account
for propagation through the Galaxy, we set upper bounds on the
annihilation cross section via a comparison with observational data.
Fig.~\ref{fig:limits-brem} collects our upper limits on the
 bremsstrahlung rate $\langle v \sigma\rangle_{\rm Brem}$.  
Although $\mu=(M_\eta/M_\chi)^2=1.2$ was chosen for display, our
conclusions remain valid for any value of $\mu\alt 6$ where the
bremsstrahlung processes dominate the $2\rarr 2$~body processes.  

\begin{figure*}[ht]
\includegraphics[height=0.23\textheight,width=0.40\columnwidth]{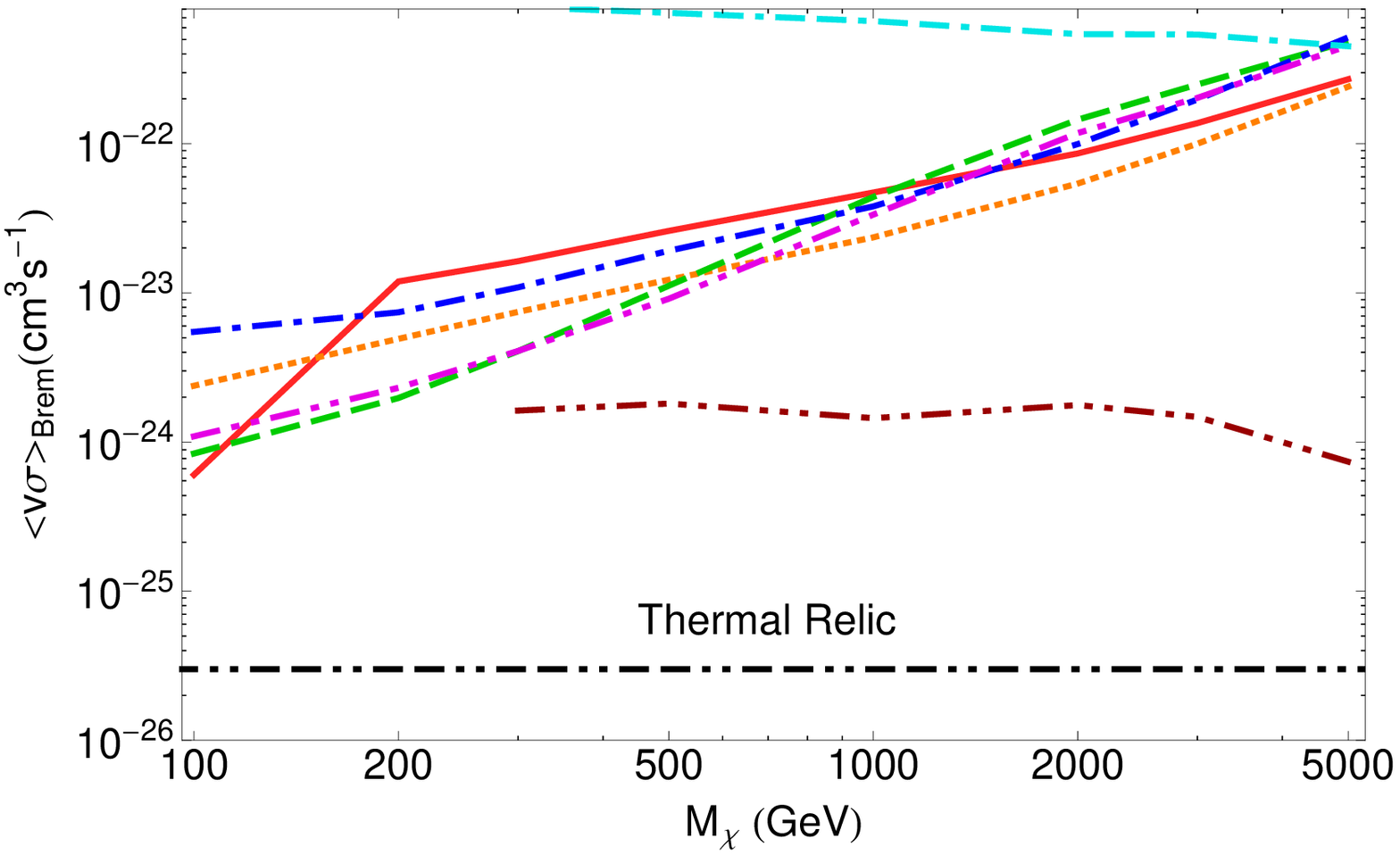}
\includegraphics[height=0.23\textheight,width=0.40\columnwidth]{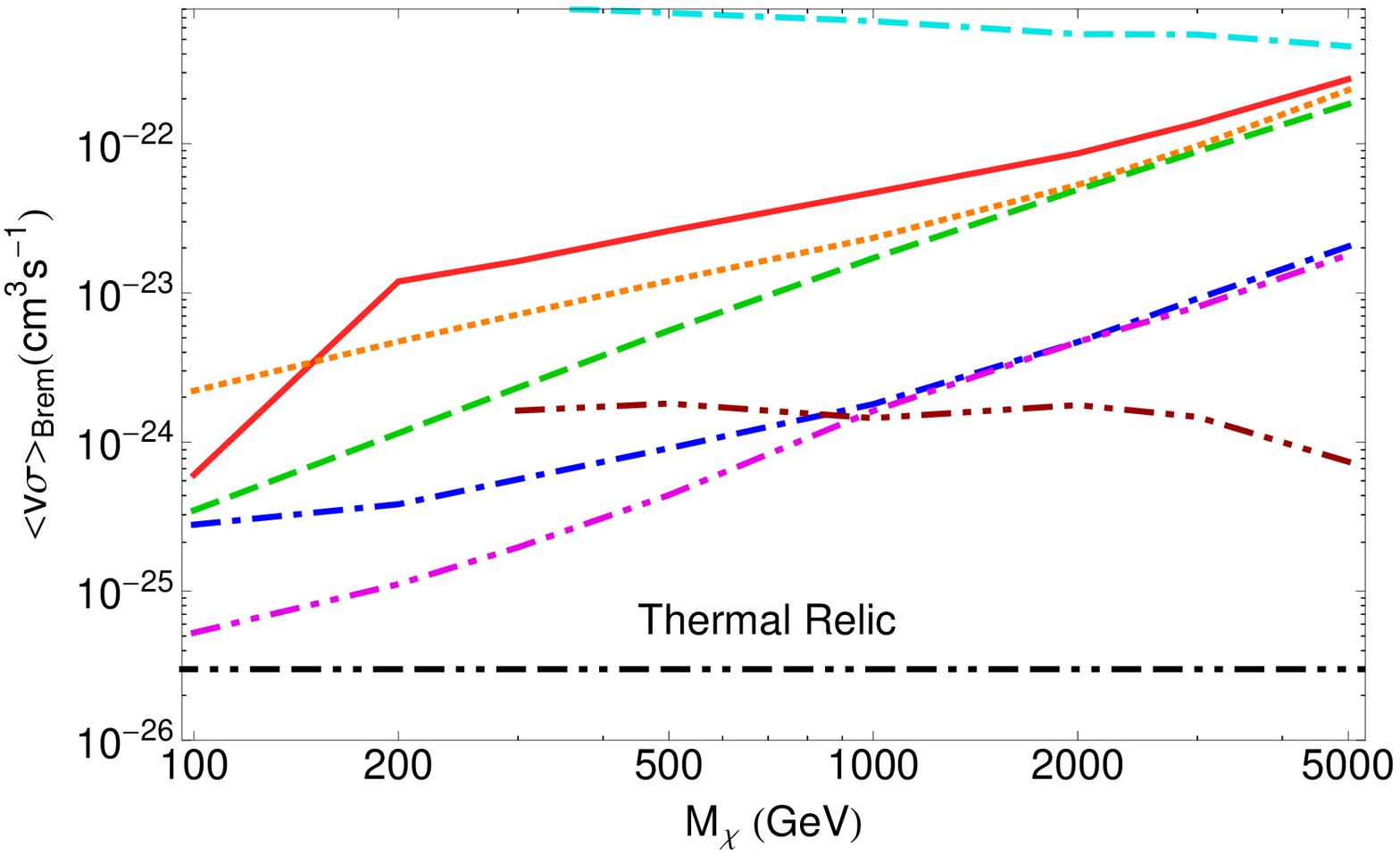}
\caption{As for Fig.~\ref{fig:limits-brem}, but using the ``min'' (left) and ``max'' (right) diffusion parameter  sets.
\label{fig:limits-comparison}}
\end{figure*}

While our analysis techniques are conservative, there are large
astrophysical uncertainties in the propagation of charged particles
through galactic magnetic fields, and in the DM density profile
which probably contains substructure.  Consequently, our
constraints are illustrative of the upper limit on the cross section, but not robust.
We assumed astrophysical propagation parameters~\cite{Cirelli:2008id} which are consistent
with a `median' antiproton flux~\cite{Donato:2003}.  However, by assuming
alternate parameters, e.g. from the `max' or `min' antiproton flux scenarios, our
results may be strengthened or weakened by up to an order of magnitude,
as shown in Fig.~\ref{fig:limits-comparison}.
We found that our conclusions hold in all cases considered except for the extreme
choice of  the ``min'' diffusion parameter set.

For the ``med'' parameter set, the constraint from the antiproton ratio is stronger than that from the
positron data by a factor of $\sim 5$.
Therefore, if the observed positron fraction were
attributed to the bremsstrahlung process, then the same process would
overproduce antiprotons by about a factor of five.
Conversely, if the bremsstrahlung process saturates
the allowed antiproton limit, then the same process produces positrons
at a rate down from the observed value by about a factor of five.  
Thus, our stringent cosmic ray antiproton limits preclude a sizable
DM contribution to observed cosmic ray positron fluxes in the
class of models for which the $W$/$Z$-bremsstrahlung processes dominate.

It is important to note that the observed antiproton flux and ratio
are well reproduced by standard astrophysical processes, leaving
little room for a DM contribution~\cite{Serpico:2011wg}.
So constraints from antiprotons are likely to be even stronger than presented here.

\section{Conclusions}
If DM is Majorana in nature, then its annihilation to SM 
fermions is suppressed due to helicity considerations.  However,
both electroweak and photon bremsstrahlung lift this suppression ,
thereby becoming the dominant channels for DM annihilation
(EW exceeding EM if the DM mass exceeds $\sim 150$~GeV,
according to the third panel in Fig.~\ref{fig:rates}).  
In this talk, we have explored the $2\rarr 2$ helicity suppression of the $s$-wave, 
and the $2\rarr 3$ un-suppression of the $s$-wave in some detail.

Unsuppressed production and subsequent decay of the emitted $W$ and $Z$ gauge bosons will produce fluxes of hadrons,
including antiprotons, in addition to electrons, positrons, neutrinos,  and gamma rays. 
This may permit indirect detection of model DM for which the annihilation rate
would otherwise be too suppressed to be of interest.

Importantly, we find that the observational data pertaining
to the flux of antiprotons make it
difficult for helicity-suppressed $2\rarr 2$ 
leptophilic DM annihilation to be the source of
the recently detected cosmic positron anomalies.
The primary culprit is the hadronic decays of the EW gauge bosons, 
which leads to a significant antiproton flux.


\vspace{0.5cm}

Acknowledgements:
I wish to acknowledge my collaborators on this $s$-wave/$p$-wave analysis of annihilation modes of 
Majorana dark matter, Nicole Bell, James Dent, and Thomas Jacques, originally, and  
Ahmad Galea and Lawrence Krauss as well, ultimately.
I thank Danny Marfatia for encouraging me to writeup this talk, and 
I thank the Center for Theoretical Underground Physics (CETUP* 2012) in South Dakota 
for hospitality and partial support during the Dark Matter Workshop where this talk was presented.
Research discussed herein was partially supported by U.S. DOE award DE-FG05-85ER40226.


\begin{thebibliography}{9}
\bibitem{note1}
The M\"oller velocity is the generalization of the relative velocity 
(as seen from either initial particle's rest frame) that makes the time-integrated rate a relativistic invariant number.
Explicitly, the M\"oller velocity ${\rm v_{\rm M}}=\{ ({\vec{\rm v}}_1 - {\vec{\rm v}}_2)^2 - ({\vec{\rm v}}_1\times {\vec{\rm v}}_2)^2 \}^\half
=  \{ ({\rm v}_1 - {\rm v}_2)^2 + {\rm v}_1 {\rm v}_2( 1-\cos\theta_{12})^2 \}^\half$
where the initial particle velocities are defined in any frame.
(The invariant is $ 2E_1 E_2 {\rm v}_{\rm M} = \sqrt{\lambda(s,M_1^2,M_2^2)}$,
where $\lambda$ is the triangle function.)
Although defined in terms of physical velocities, the M\"oller velocity is not itself a physical velocity;
in fact, it can exceed unity.
Commonly used limits result (i) in the rest frame of one initial particle, say ${\rm v}_2=0$,
where ${\rm v_{\rm M}}={\rm v}_1={\rm v}_{\rm rel}$, as it must; 
and (ii) when both initial particles are relativistic,
where ${\rm v_{\rm M}}=1-\cos\theta_{12}$.
An informative discussion and derivation of ${\rm v}_{\rm M}$ is given in~\cite{LL}, 
and nuances of thermal averaging with ${\rm v}_{\rm M}$ are given in~\cite{GelmGondo}.
\bibitem{note2}
One mnemonic for the general result is that the $L^{th}$ partial wave includes a factor  
$P_L(\cos\bar\theta)$, 
and since the amplitude is an analytic function of Lorentz invariants $s$, $t(\cos\bar\theta)$, and $u(\cos\bar\theta)$,  
the true argument of $P_L$ is $\propto {\bar p}_{\rm i} {\bar p}_{\rm f} \cos\bar\theta$;
the latter scales as ${\bar {\rm v}}_{\rm i} {\bar{\rm v}}_{\rm f}$.
Integration over singularities picks out the highest power of $P_L$, and we get
an amplitude scaling as $({\bar {\rm v}}_{\rm i} {\bar{\rm v}}_{\rm f})^L$.
(With spins present, $L$ is replaced by $J$ 
and $P_L(\cos\bar\theta)$ by the Wigner function $d^J_{\mu\lambda}(\bar\theta)$.
But these replacements are immaterial to the mnemonic, as any mnemonic is just a mnemonic.)
\\
Another mnemonic is to express with the available vectors, ${\vec p}_{i}$ and ${\vec p}_{f}$, 
the fact that $L$ tells us how the system transforms under spatial rotations. 
The unique  characterization for a system at fixed $L$ is the $L^{th}$ power of ${\vec p_{i}} \cdot {\vec p_{f}}$ in the CoM,
which when squared and  integrated over $\cos\bar\theta$  gives a rate proportional to $({\bar{\rm v}}_{i} {\bar{\rm v}}_{f})^{2L}$.
\end{thebibliography}
\end{document}